\documentclass[twocolumn]{aastex63}
\usepackage[utf8]{inputenc}
\usepackage{graphicx}
\usepackage{color}

\DeclareGraphicsExtensions{.ps,.eps,.pdf,.jpg,.png}

\newcommand{\kms}{\ifmmode{\rm km\thinspace s^{-1}}\else km\thinspace s$^{-1}$\fi}
\newcommand{\rsun}{R$_\odot$}
\newcommand{\msun}{M$_\odot$}
\newcommand{\eb}{2M1222$-$57}
\newcommand{\teff}{\ensuremath{T_{\rm eff}}}

\begin{document}

%\maketitle

\title{A Low-Mass Pre--Main-Sequence Eclipsing Binary in Lower Centaurus Crux Discovered with TESS}
\author[0000-0002-3481-9052]{Keivan G.\ Stassun}
\affiliation{Department of Physics and Astronomy, Vanderbilt University, Nashville, TN 37235, USA}
\author[0000-0002-5286-0251]{Guillermo Torres}
\affiliation{Center for Astrophysics $\vert$ Harvard \& Smithsonian, 60 Garden Street, Cambridge, MA 02138, USA}
\author[0000-0002-5365-1267]{Marina Kounkel}
\affiliation{Department of Physics and Astronomy, Vanderbilt University, Nashville, TN 37235, USA}
\author[0000-0002-2457-7889]{Dax L.\ Feliz}
\affiliation{Department of Physics and Astronomy, Vanderbilt University, Nashville, TN 37235, USA}
\author[0000-0002-0514-5538]{Luke G.\ Bouma}
\affiliation{Cahill Center for Astrophysics, California Institute of Technology, Pasadena, CA 91125, USA}

\author[0000-0002-2532-2853]{Steve~B.~Howell}
\affiliation{NASA Ames Research Center, Moffett Field, CA 94035, USA}
\author[0000-0003-2519-6161]{Crystal~L.~Gnilka}
\affiliation{NASA Ames Research Center, Moffett Field, CA 94035, USA}
\author[0000-0001-9800-6248]{E. Furlan}
\affiliation{NASA Exoplanet Science Institute, Caltech/IPAC, Mail Code 100-22, 1200 E. California Blvd., Pasadena, CA 91125, USA}

\begin{abstract}
We report the discovery of \eb\ as a low-mass, pre--main-sequence (PMS) eclipsing binary (EB) in the Lower Centaurus Crux (LCC) association for which, using {\it Gaia\/} parallaxes and proper motions with a neural-net age estimator, we determine an age of 16.2$\pm$2.2~Myr. 
The broadband spectral energy distribution (SED) shows clear excess at $\gtrsim$10~$\mu$m indicative of a circumbinary disk, and new speckle-imaging observations reveal a faint, tertiary companion separated by $\sim$100~AU.
H$\alpha$ emission is modulated on the orbital period, consistent with theoretical models of orbitally pulsed accretion streams reaching from the inner disk edge to the central stars.
From a joint analysis of spectroscopically determined radial velocities and TESS light curves, together with additional tight constraints provided by the SED and the {\it Gaia\/} parallax, we measure masses for the eclipsing stars of 0.74~\msun\ and 0.67~\msun; radii of 0.98~\rsun\ and 0.94~\rsun; and effective temperatures of 3750~K and 3645~K. The masses and radii of both stars are measured to an accuracy of $\sim$1\%. 
The measured radii are inflated, and the temperatures suppressed, relative to predictions of standard PMS evolutionary models at the age of LCC; also, the Li abundances are $\sim$2~dex less depleted than predicted by those models. However, models that account for the global and internal effects of surface magnetic fields are able to simultaneously reproduce the measured radii, temperatures, and Li abundances at an age of $17.0 \pm 0.5$~Myr. 
Altogether, the \eb\ system presents very strong evidence that magnetic activity in young stars alters both their global properties and the physics of their interiors.
\\
\end{abstract}

\section{Introduction}

Eclipsing binary (EB) star systems have long served as laboratories for the most precise measurements of fundamental stellar parameters. Indeed, the best characterized EBs can produce measurements of stellar
masses and radii that are precise and accurate to the percent level \citep[e.g.,][]{Torres:2010}. When compared against
grids of stellar models or isochrones, the predictions of stellar theory can be stringently tested. In this way, key input parameters in stellar models can be refined and missing physical ingredients in the models can be identified and empirically constrained. 

Such tests of stellar models are especially important in the pre--main-sequence (PMS) stage of evolution, where multiple physical effects are known to operate that stress the simpler assumptions of theory for main-sequence stars. 
For example, low-mass PMS stars (i.e., T~Tauri stars) often exhibit phenomena attributed to strong surface magnetic fields---such as rapid rotation, chromospheric and coronal activity, accretion from surrounding protoplanetary disk material---and there is evidence that these effects can alter the bulk properties and internal structures of the stars \citep[see, e.g.,][for a review]{Stassun:2014}. 

As specific case studies, consider the PMS EBs V1174~Ori \citep[two roughly solar-mass stars, age of $\sim$10~Myr;][]{Stassun:2004}, Par~1802 \citep[two identical 0.4~\msun\ stars, age of $\sim$1~Myr;][]{Stassun:2008} and 2M0535$-$05 \citep[two brown dwarfs, age of $\sim$1~Myr;][]{Stassun:2006,Stassun:2007}. In the case of V1174~Ori, the unexpectedly small amount of Li depletion observed implies significantly suppressed surface convection, as might result from strong surface fields. In the case of Par~1802, the two stars have radii that differ by $\sim$10\% and temperatures that differ by $\sim$10\%---and thus luminosities that differ by $\sim$60\%---despite having masses that are identical to $\sim$2\%, possibly the result of differing magnetic field strengths and/or interactions with a tertiary companion \citep[see also][]{Gomez:2012}. 2M0535$-$05 may be the most dramatic case, in which there is a surprising reversal of temperatures with mass, such that the higher-mass brown dwarf is actually cooler than its lower-mass companion. Followup studies have linked this ``temperature suppression" effect to magnetic chromospheric activity, and have suggested that a ``radius inflation" effect accompanies it \citep[see, e.g.,][]{Stassun:2012,Somers:2017,Jaehnig:2019}. 
Other recent case studies of these effects in low-mass PMS EBs include, e.g., \citet{David:2019}, \citet{Murphy:2020}, \citet{Tofflemire:2022}, and references therein, as well as additional references in the review by \citet{Stassun:2014}.
In summary, there is now strong evidence that PMS evolutionary models need to incorporate the physical effects of strong surface magnetic fields to explain the observations, and PMS EBs have been crucial for the precise measurements leading to these insights. 

Thanks to the availability of precise photometry and astrometry from all-sky surveys, the ability of EBs to serve as astrophysical benchmarks for fundamental stellar properties has been dramatically increased in recent years. In particular, with broadband fluxes spanning far-ultraviolet to mid-infrared wavelengths for millions of stars across the sky, from which full spectral energy distributions (SEDs) and precise bolometric fluxes can be measured, the best characterized EBs have been shown to serve as highly accurate standard candles \citep[e.g.,][]{StassunTorres:2016} capable of testing trigonometric parallaxes down to the mas level \citep[e.g.,][]{StassunTorres:2021}. 

In addition, with the advent of high-precision parallaxes provided by {\it Gaia\/} \citep[see, e.g.,][]{Brown:2021}, recent work has demonstrated that EBs can also serve as benchmarks in ways that were not previously possible, including the ability to provide tight constraints on effective temperature (\teff) that are complementary to the constraints provided by the eclipsing nature of the system. For example, \citet{Miller:2020} have shown that, with the availability of a well characterized SED together with a precise parallax, the \teff\ of each star in an EB can be constrained to $\sim$0.2\% (i.e., $\sim$15~K for a solar-type star)---vastly better than the traditional limitation of systematic error floors on \teff\ of $\sim$100~K---representing a tremendous advance in the ability of EBs to stringently test stellar models. 

{\it Gaia\/} has moreover revolutionized the ability to associate stars with clusters and groups across the Galaxy, vastly extending the traditional identification of massive clusters or very nearby, high proper-motion moving groups. This is important because stellar clusters and associations offer the opportunity to test and refine theories of stellar structure and evolution at the population level; given their common distance, chemical composition, and assumed common age, clusters and associations act as laboratories to calibrate models and physical mechanisms that must be able to simultaneously explain the observed characteristics of cluster or group members at a common, independently determined age.  

For example, \citet{Kounkel:2020} have identified thousands of clusters, associations, and groups out to distances as far as 3~kpc and ages up to 1~Gyr, with numbers of stellar members ranging from tens to thousands. Many of these groups, especially at ages younger than 100~Myr, span very large swaths of the sky in both angular and physical extent, and thus would have been very difficult to identify prior to {\it Gaia}. This work has also helped to refine the boundaries and memberships of previously known associations and groups in the solar neighborhood. 
Importantly, these and other studies \citep[e.g.,][]{Kounkel:2019,McBride:2021} have shown that it is possible to assign precise, semi-empirical ages to individual members of these groups and associations. Thus, an EB that can be shown to be a member of one of these groups or associations has the added advantage of an independently determined age that can help to further stress-test the physics of PMS evolutionary models. 

In this paper, we report the discovery of \eb\ as a PMS M+M EB and perform a detailed investigation of its physical properties in comparison to the predictions of PMS stellar evolution models. 
First, in Section~\ref{sec:tic411} we use the parallax and space motions from {\it Gaia\/} 
DR3 \citep{Lindegren:2018,Luri:2018} 
%together with spectroscopic observations of Lithium absorption and H$\alpha$ emission 
to investigate the membership of this system in the young Lower Centaurus Crux association, and we use the methods of \citet{Kounkel:2020} and \citet{McBride:2021} to assign a precise age based on that association.
Next, using new space-based 
photometry from the Transiting Exoplanet Survey Satellite \citep[TESS,][]{Ricker:2015} combined with newly obtained radial velocities and broadband photometry from the literature (Section~\ref{sec:data}), we perform 
eclipse and SED modeling of the system to precisely determine the fundamental stellar properties of the system (Section~\ref{sec:results}), and we furthermore investigate residuals in the light curve for 
rotational and disk-accretion signatures.  Finally, 
we discuss this system in an evolutionary context in Section~\ref{sec:discussion}, including tests of PMS stellar evolution models, evidence for the effects of magnetic fields on the temperatures and radii of low-mass stars, as well as the role of circumbinary disk interactions and accretion streams to explain the observed out-of-eclipse variations. We conclude with a summary of our findings in Section~\ref{sec:summary}.

\begin{figure*}[!ht]
    \centering
    \includegraphics[width=\linewidth,trim=0 0 0 20,clip]{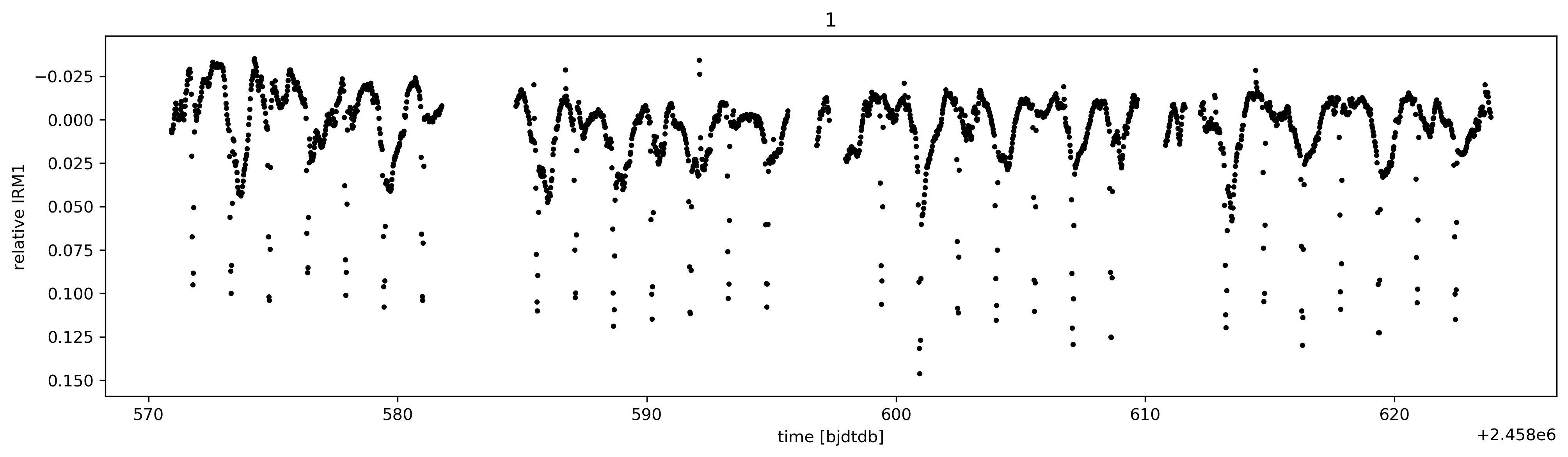}
    \includegraphics[width=\linewidth,trim=0 0 0 48,clip]{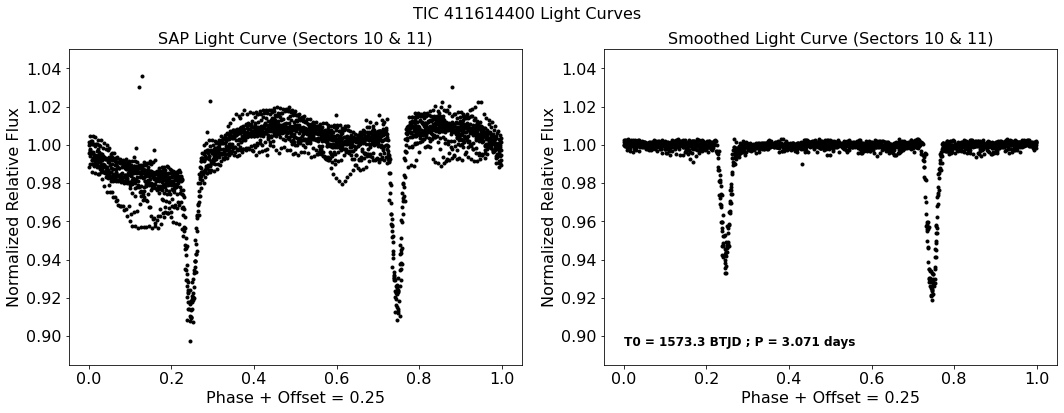}
    \caption{(Top) Discovery light curve of \eb\ from TESS FFI observations of sectors 10 and 11 \citep[see][]{Bouma:2019}. (Bottom) Light curve phase folded on period of 3.071~d with (right) and without (left) out-of-eclipse variations removed. Note that the out-of-eclipse variations include both a periodic modulation with peak-to-peak amplitude of $\sim$0.075~mag and an additional, more stochastic, source of variability at the level of $\sim$1\%.}
    \label{fig:cdips_lc}
\end{figure*}

\section{The \texorpdfstring{\eb}{2M1222--57} Eclipsing Binary System: Variability, Group Membership, Age, and Distance Considerations}\label{sec:tic411} 

\subsection{Discovery Light Curve Variability Characteristics}\label{subsec:discovery}

\eb\ was identified as a likely pre--main-sequence (PMS) eclipsing binary (EB) in TESS light curve data as part of the Cluster Difference Imaging Photometric Survey \citep[CDIPS; see][]{Bouma:2019} of young star-forming regions. Its identifier in the TESS Input Catalog \citep[TIC;][]{StassunTIC:2019} is TIC 411614400, and the TIC cross-matches it in various other source catalogs as 2MASS J12220147$-$5737565, WISEA J122201.42$-$573756.6, ASAS J122201$-$5737.8, ROSAT 2RXS J122201.0$-$573754, UCAC4 162$-$087608, among others. Its {\it Gaia\/} DR2 identifier is 6071734350857944704. We opt to refer to the object by its shorthand 2MASS identifier: \eb. 

The CDIPS light curve from the Full Frame Images \citep[FFIs; see, e.g.,][]{Oelkers:2018} in TESS observing sectors 10 and 11 (Figure~\ref{fig:cdips_lc}, top) exhibited clearly repeating, punctuated flux dropouts with depths of $\sim$0.15~mag superposed on a periodic modulation with peak-to-peak amplitude of $\sim$0.075~mag. The periodic modulation appears to have roughly the same period as the flux dropouts, but the shape of the modulation varies somewhat from cycle to cycle (Figure~\ref{fig:cdips_lc}, bottom left), suggesting the presence of some additional source of variability at the level of $\sim$1\% amplitude that is more stochastic in nature. There were also a few possible flaring events observed. Applying a simple spline fit to the out-of-eclipse variations reveals a clearly well detached EB light curve with period of 3.071~d (Figure~\ref{fig:cdips_lc}, bottom right). The primary and secondary eclipses are separated by 0.5 phase and have depths of $\sim$8\% and $\sim$6\%, respectively, suggesting a circular orbit and a grazing inclination angle. 

\subsection{Group Membership, Age, and Distance}\label{subsec:age}

\begin{figure}[!t]
    \centering
    \includegraphics[width=\linewidth]{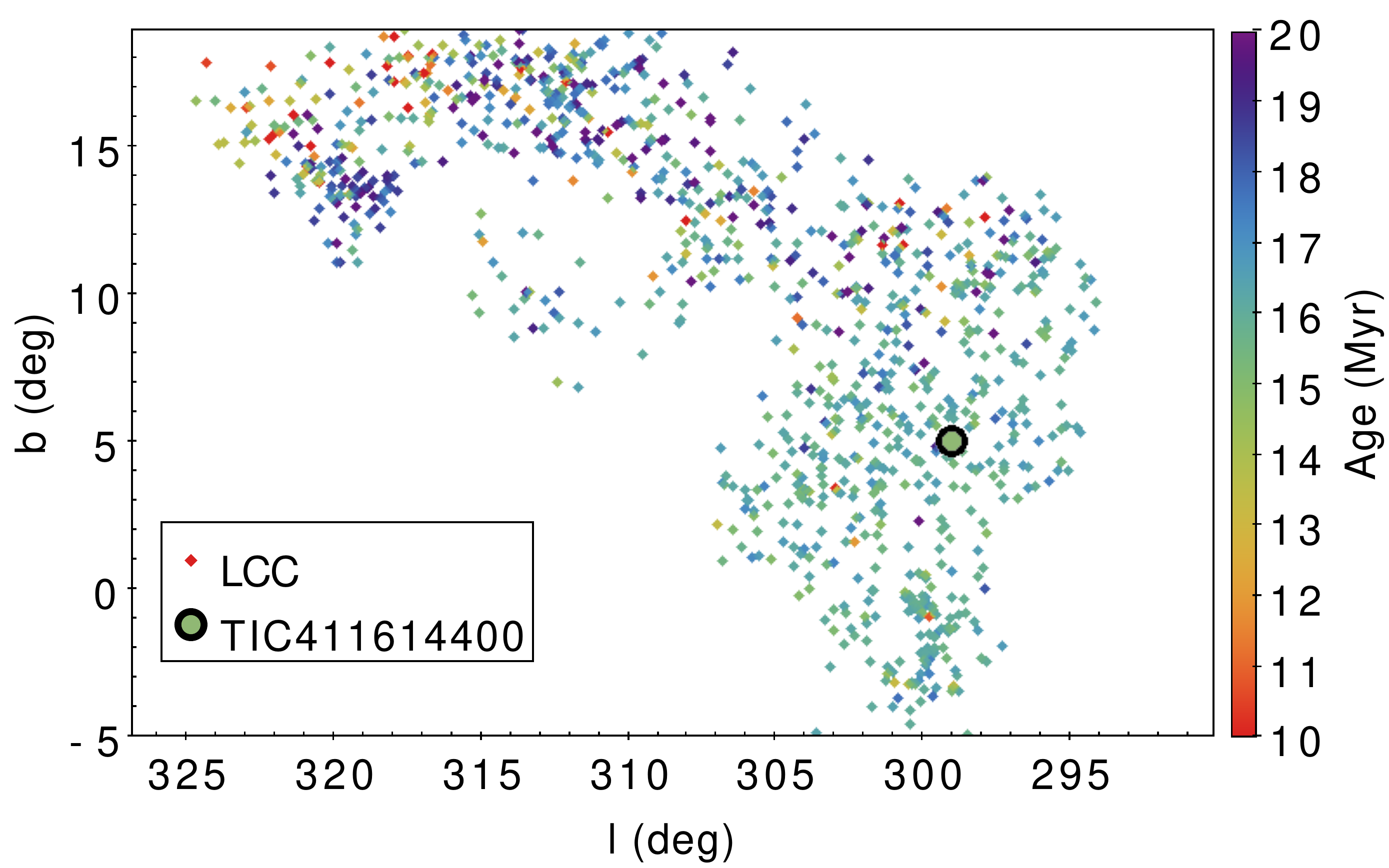}
    \includegraphics[width=\linewidth]{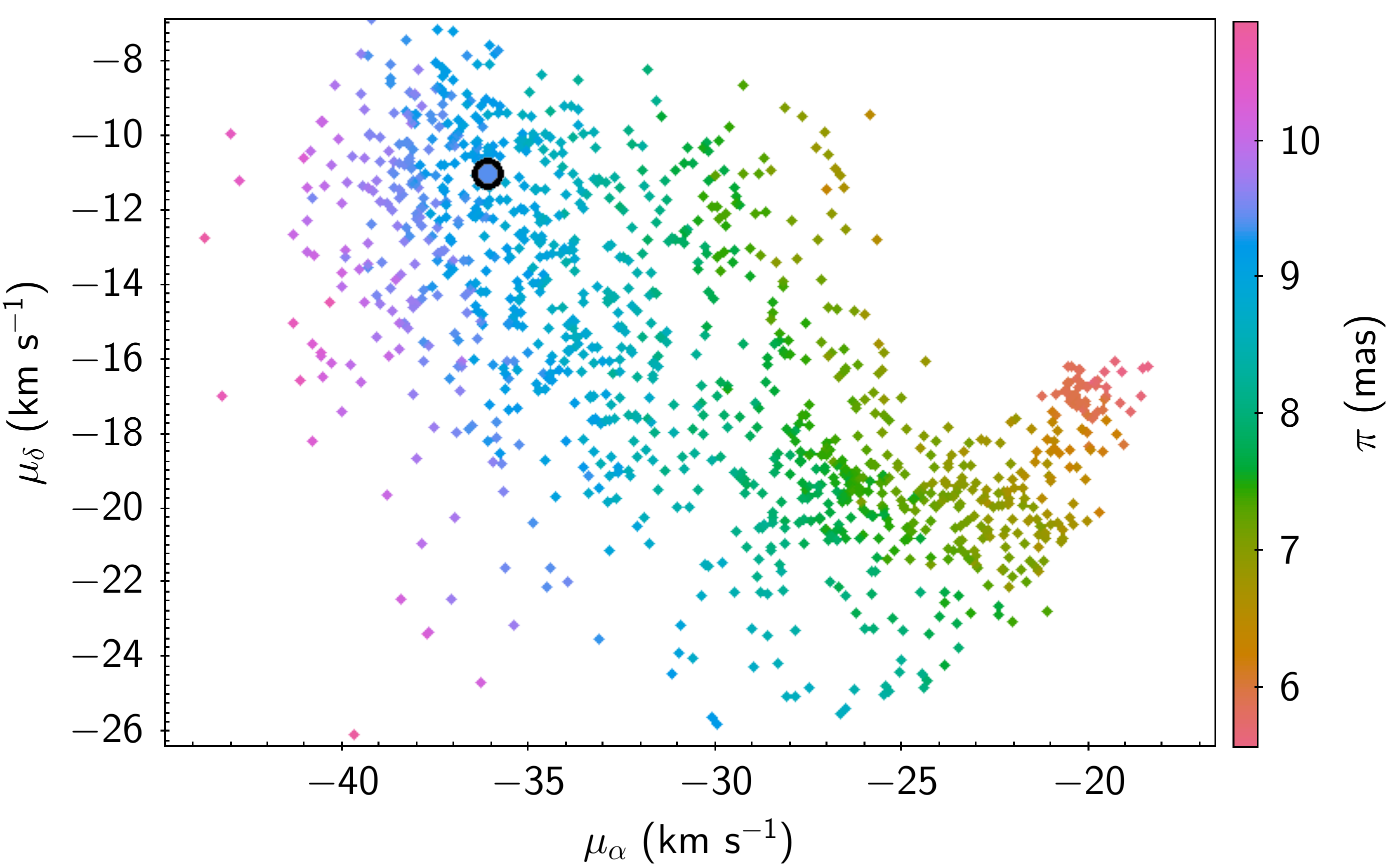}
    \caption{Spatial map of LCC and its proper motions \citep[from][]{McBride:2021} color-coded by the inferred age and parallax respectively. \eb~ is highlighted against LCC with a larger circle with black outline.}
    \label{fig:agemap}
\end{figure}

\eb\ has been identified by previous surveys as a PMS star on the basis of the presence of H$\alpha$ emission and very strong Li absorption \citep[see][]{Bowler:2019}. In addition, several studies of the spatial and kinematic distributions of young stars in the region have associated \eb\ with young moving groups or other coherent, young stellar structures in the solar neighborhood. For example, \citet{Kounkel:2020} identify \eb\ to be part of the Sco Cen OB association (Figure~\ref{fig:agemap}), namely towards Lower Centaurus Crux (LCC). 

Using Sagitta, a neural net based algorithm that enables age estimates of PMS stars from their astrometry and photometry \citep{McBride:2021}, we derive an age of $\log\tau = 7.21 \pm 0.06$ ($16.2 \pm 2.2$~Myr). This is consistent with the age of $\sim$16~Myr inferred for LCC by several other studies  \citep[see, e.g.,][and references therein]{Preibisch:2008}. 

\begin{figure}[!ht]
    \centering
    \includegraphics[width=\linewidth]{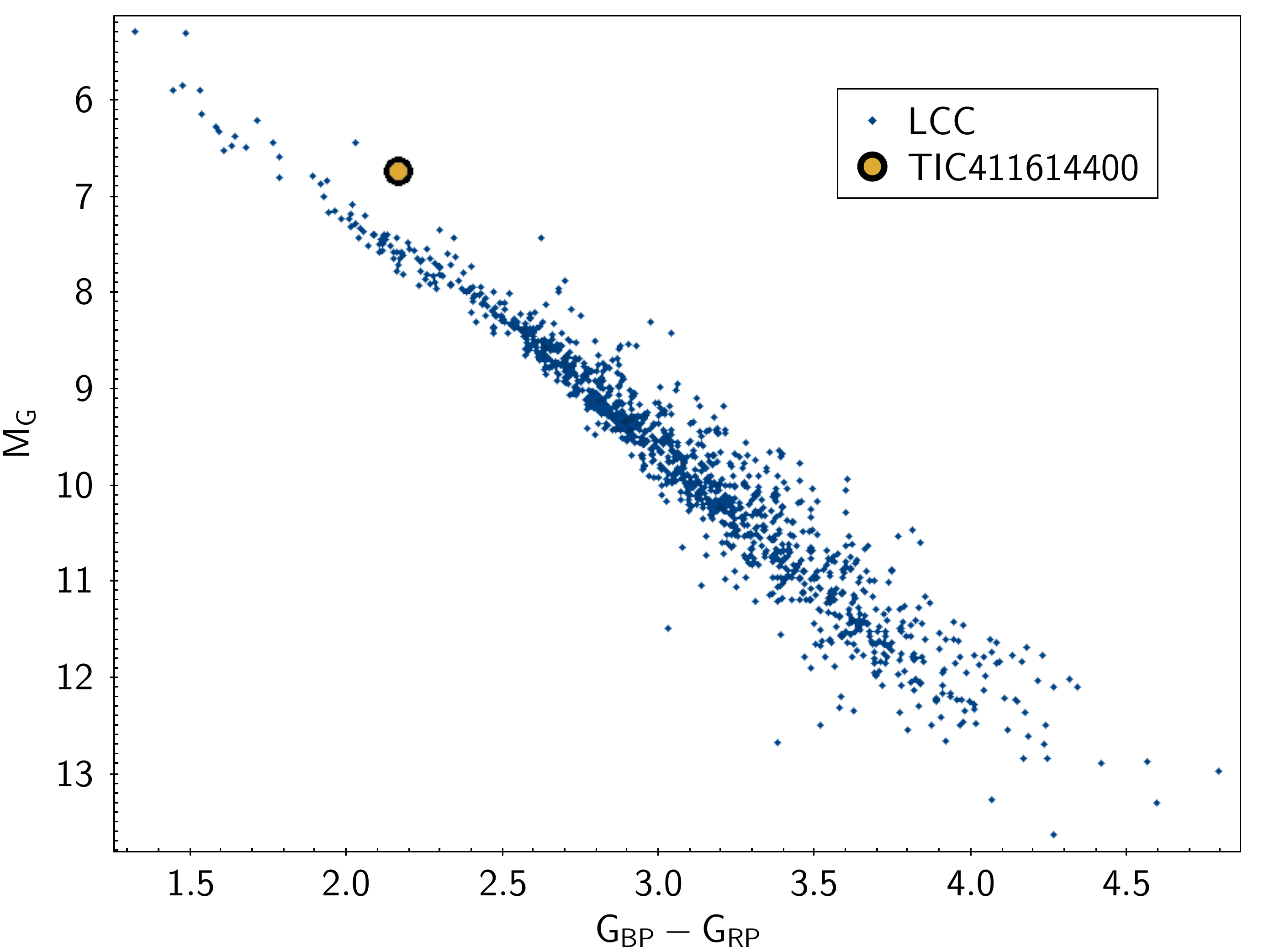}
    \caption{{\it Gaia\/} color-magnitude diagram of LCC \citep[from][]{McBride:2021} in black, and \eb\ marked by the yellow circle.}
    \label{fig:cmd}
\end{figure}

\eb\ stands out among the other members of LCC, its position in the color-magnitude diagram distinguishing it as one of the most luminous objects for its color (Figure~\ref{fig:cmd}). This is attributable to its placement on the cluster binary sequence rather than being younger than its siblings: \citet{Bowler:2019} identify it as a likely spectroscopic binary with spectral type of approximately M1. The {\it Gaia\/} reported RUWE value of 1.5 is also consistent with the presence of photocenter motion as would be expected for a binary with stars of near-equal brightness \citep[see, e.g.,][]{StassunTorres:2021}. 

Finally, the {\it Gaia\/} DR3 parallax measurement of $9.453 \pm 0.025$~mas \citep[including the most recent published global parallax offset and inflation factor on the parallax uncertainty; see][]{Lindegren:2021, ElBadry:2021} securely places \eb\ at a distance of $105.79 \pm 0.28$~pc. As we will see below, this precise distance measurement helps us place stringent constraints on the temperatures and radii of the eclipsing stellar components of the \eb\ system. 

\section{Data}\label{sec:data}

\subsection{TESS Light Curve}\label{subsec:photometry} 

\begin{figure*}[!ht]
    \centering
    \includegraphics[width=\linewidth,trim=0 0 0 260,clip]{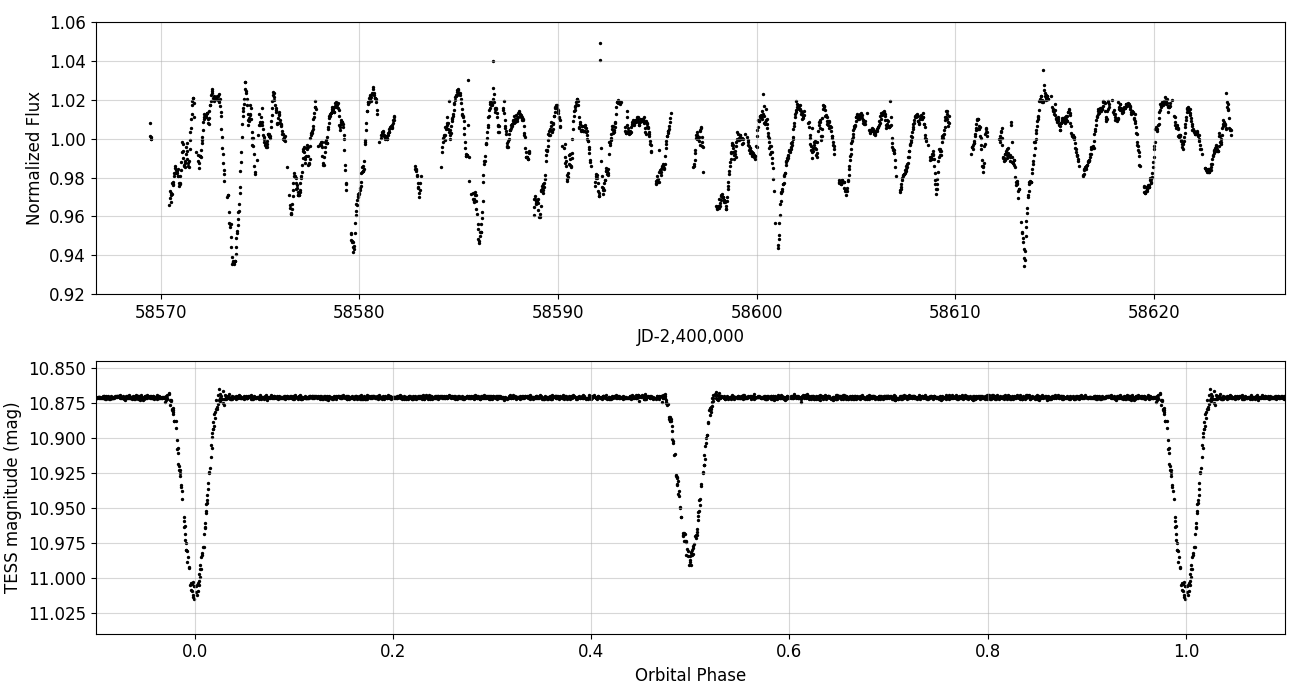}
    \caption{%(Top) TESS FFI light curve from Sectors 10--11, with points in eclipse removed to isolate the out-of-eclipse variations. (Bottom) 
    Phase-folded TESS FFI light curve from Sectors 10--11 (Figure~\ref{fig:cdips_lc})
    %light curve of the primary and secondary eclipses 
    after smoothing out the out-of-eclipse variations.}
    \label{fig:TESS_lc}
\end{figure*}

\begin{figure*}[!ht]
    \centering
    \includegraphics[width=\linewidth,trim=0 55 0 0,clip]{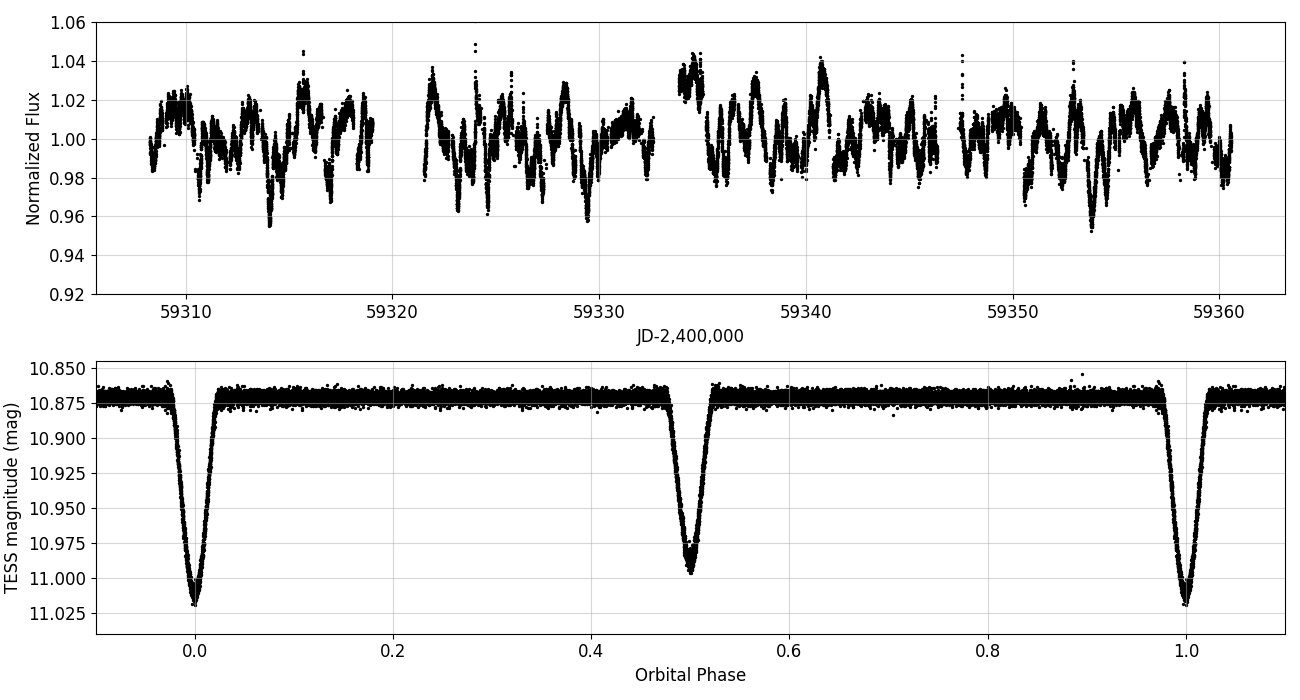}
    \includegraphics[width=\linewidth,trim=0 0 0 260,clip]{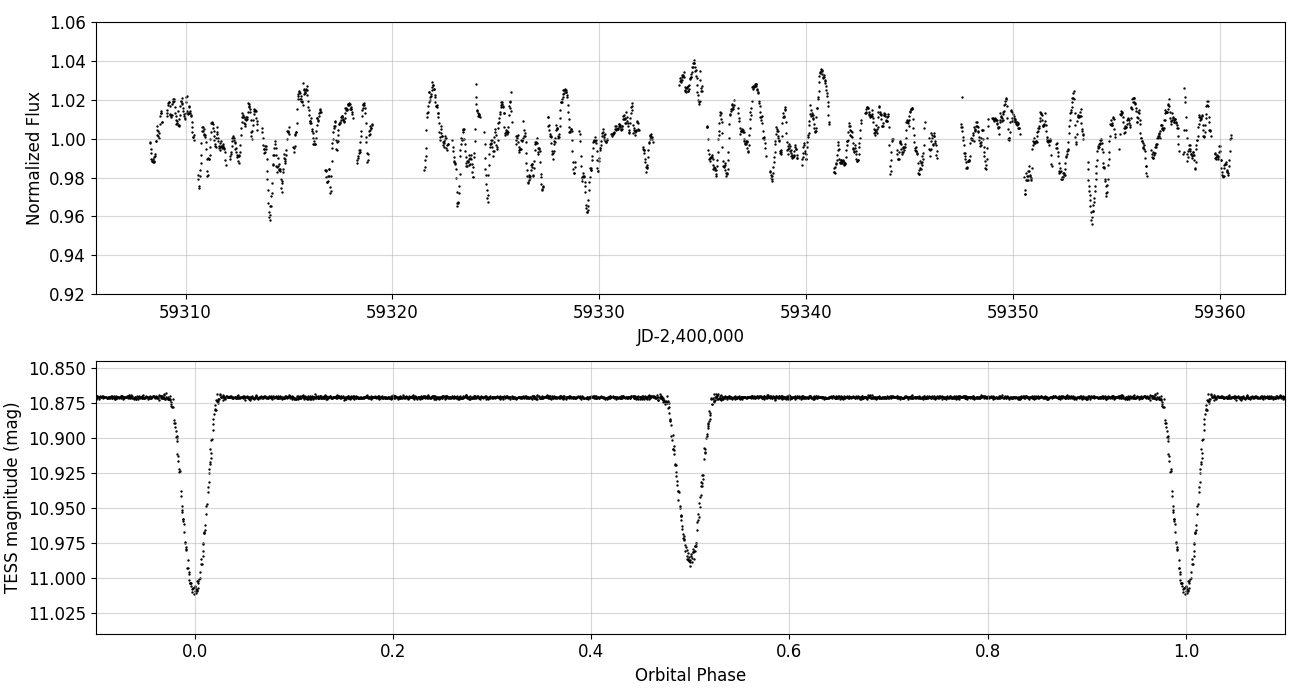}
    \caption{%Same as Figure~\ref{fig:TESS_lc}, but here showing the 
    TESS 2-min cadence light curve from Sectors 37--38. (Top) Points in eclipse have been removed to isolate the out-of-eclipse variations. (Middle) Phase-folded on the orbital period after smoothing out the out-of-eclipse variations. (Bottom) Same as middle, but binned by 30~min for comparison with Sectors~10--11 (Figure~\ref{fig:TESS_lc}).}
    \label{fig:TESS_lc_2}
\end{figure*}

TESS observed \eb\ in its 30-min FFI mode over nearly 60~d in Sectors 10--11, and again in its 2-min mode over another nearly 60~d in Sectors 37--38 (Figures~\ref{fig:cdips_lc}, \ref{fig:TESS_lc}--\ref{fig:TESS_lc_2}).  The latter observations were through the Director’s Discretionary Target program (PI: Bouma).
%The pixel mask used to extract the light curve from the TESS FFIs is shown in Figure~\ref{fig:TESS_mask}.  

\begin{figure*}[!ht]
    \centering
    \includegraphics[width=\linewidth,trim=0 0 0 0,clip]{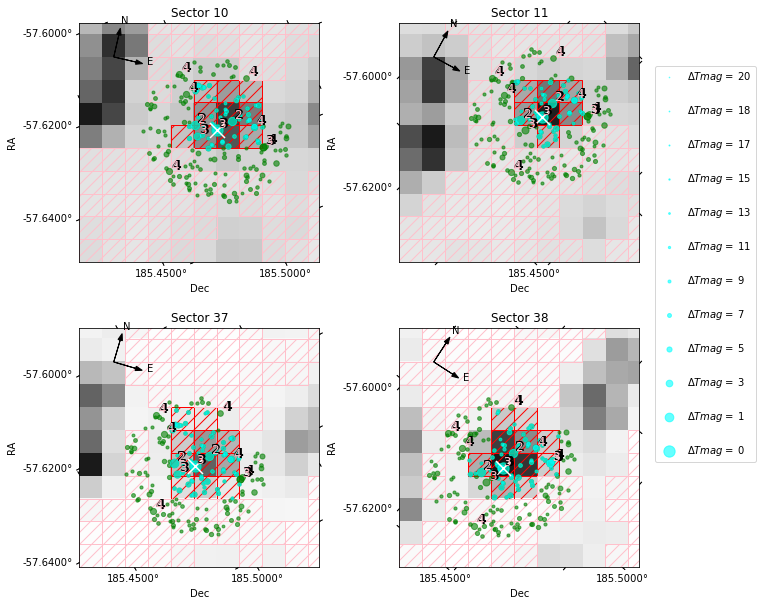}
    \caption{Cutouts of Full Frame Images from TESS Sectors 10--11, and of 2-min observations from TESS Sectors 37--38, centered on \eb. Other stars from the TIC within (cyan) and outside of (green) the pixel mask are shown along with their apparent TESS magnitudes.}
    \label{fig:TESS_mask}
\end{figure*}

Unfortunately there is no flux contamination estimate provided by the TIC because this star was not selected for inclusion in the TESS Candidate Target List \citep[CTL;][]{StassunTIC:2019}.
Therefore, we constructed custom pixel masks to extract the TESS light curve from each sector based on a careful assessment of all surrounding stars included in the TIC (Figure~\ref{fig:TESS_mask}). In addition, we used the apparent TESS magnitudes of all stars in the TIC that could contribute flux within the pixel masks in order to correct for dilution of the eclipses and other variations.
We obtain flux contamination fractions (flux of contaminants relative to total flux) of 0.301, 0.276, 0.307, and 0.318, for Sectors 10, 11, 37, and 38, respectively.

%{\bf [WILLIE UPDATE THIS AS NEEDED.]}
The light curve also exhibits out-of-eclipse variations with peak-to-peak amplitude of $\sim$5\%, which include a combination of instrumental systematics and true source modulations. As shown in Figure~\ref{fig:cdips_lc}, the modulations change significantly from cycle to cycle, which complicates the analysis with an EB model. 
%We therefore opted to detrend the light curve manually, by masking out the eclipses and fitting a spline function to the out-of-eclipse portions to remove the variability. We did this separately for the data with 30\,m and 2\,m cadence. We then divided the raw light curve by these spline functions, interpolating over the eclipse sections. 
We therefore opted to detrend the light curve manually, by masking out the eclipses and fitting a cubic spline function\footnote{As implemented in the Python function {\tt scipy.interpolate.UnivariateSpline}.\label{lab:spline}} to the out-of-eclipse portions in order to remove the variability. We did this separately for the data with 30 m and 2 m cadence, and modeled both the periodic component (more clearly seen in sectors 10--11) and the more stochastic component that seems to dominate in sectors 37--38. In each case we set the smoothing factor $s$ (see footnote~\ref{lab:spline}) that controls the number of spline knots to provide a reasonable approximation as assessed visually, while at the same time avoiding overfitting of noise as well as undue variations in the masked-out eclipse sections that might affect the shape of the underlying eclipses. We then divided the raw light curve by these spline functions, interpolating over the eclipse sections.
This is the final photometry that we will use below in Section~\ref{subsec:lcfit}. The phase-folded light curves for sectors 10--11 and 37--38 are shown in the bottom panels of Figures~\ref{fig:TESS_lc} and \ref{fig:TESS_lc_2}.

\subsection{Radial Velocities}\label{subsec:rv}
\begin{figure}[!t]
    \centering
    \includegraphics[width=\linewidth]{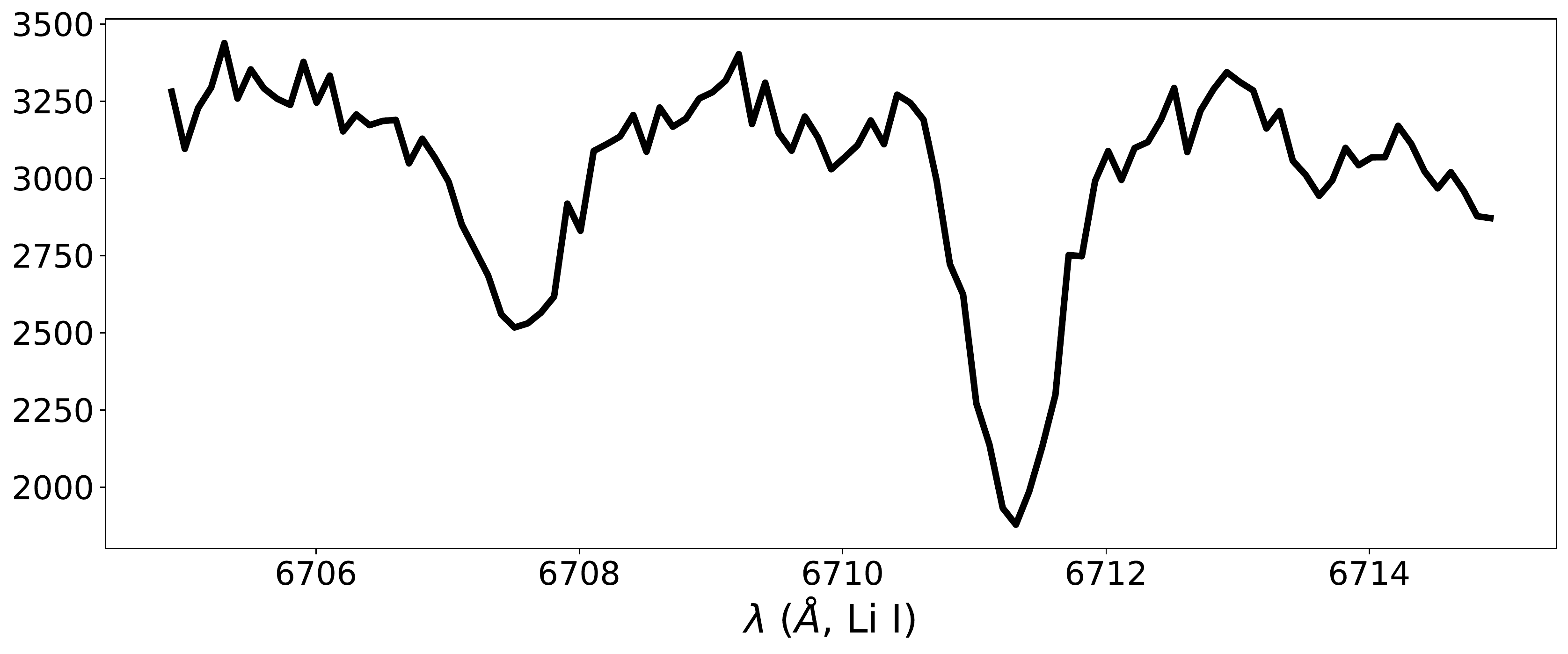}
    \includegraphics[width=\linewidth]{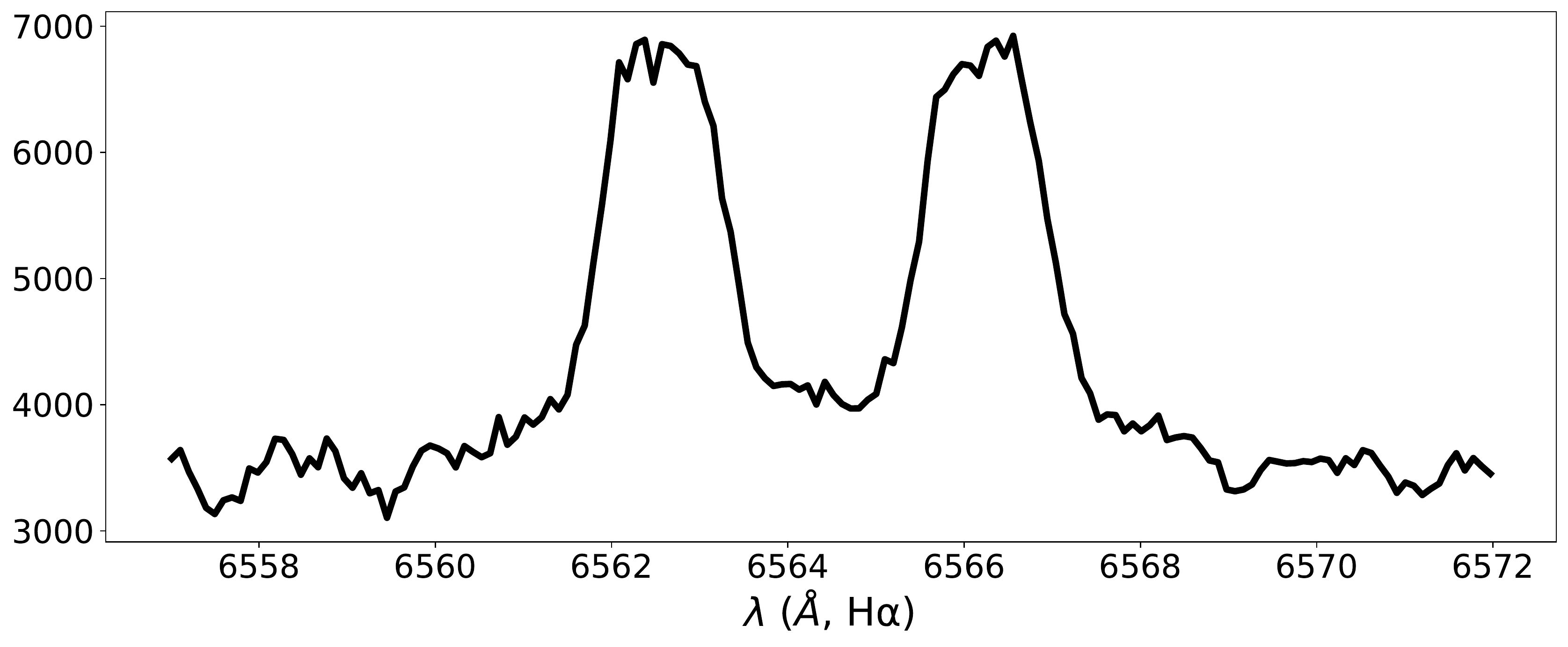}
    \caption{Representative double-line spectrum of \eb\ for the region around the Li $\lambda$6708 line (top) and the H$\alpha$ $\lambda$6563 line (bottom).}
    \label{fig:liha}
\end{figure}

\eb\ was observed at 17 separate epochs with the CHIRON echelle spectrograph on the CTIO 0.9-m telescope from 29 Jan 2021 to 28 Apr 2021
in the fiber mode configuration ($R \approx$ 25,000, $\lambda\lambda$410--870~nm).
At most epochs, the observed spectrum was clearly double-lined and the relative line strengths suggested components of similar brightness. For example, Figure~\ref{fig:liha} shows the observed spectrum at one representative epoch for the regions around the Li $\lambda$6708 line and the H$\alpha$ $\lambda$6563 line. One epoch had the two components severely blended, making it difficult to obtain accurate radial velocities; this spectrum was therefore not used for our analysis.

We extracted radial velocities (RV) for both components using the Python implementation of the IRAF XCSAO function \citep{rvsao,pyxcsao}, through performing a cross-correlation against a high resolution PHOENIX synthetic spectrum \citep{husser2013} with solar metallicity, $T_{\rm eff}=3600$~K, and $\log g$=4.0. To minimize the noise from the joining of the echelle orders, to avoid H$\alpha$ emission line that is not represented in the template, as well as to avoid strong telluric features and maximize the RV precision, cross-correlation was done only using the wavelength range of 7000--8000~\AA, and a Fourier filter was applied to minimize low spatial frequencies. Across all epochs, the cross-correlation function showed two strongly pronounced peaks, velocities of which were obtained independently through Gaussian fitting. The RVs so measured are summarized in Table~\ref{tab:rvs}.

\begin{deluxetable}{rrrrr}
\tablewidth{0pt}
\tablecaption{Radial Velocities of \eb\label{tab:rvs}}
\tablehead{
\colhead{BJD} & \colhead{RV$_1$} & \colhead{$\sigma_{\rm RV_1}$} & \colhead{RV$_2$} & \colhead{$\sigma_{\rm RV_2}$} \\
\colhead{} & \colhead{\kms} & \colhead{\kms} & \colhead{\kms} & \colhead{\kms}
}
\startdata
2459243.81852 &       88.46 &      0.34 &      $-$68.55 &      0.62 \\
2459245.73937 &      $-$25.38 &      0.53 &       56.56 &      0.69 \\
2459249.83554 &       90.93 &      0.34 &      $-$72.40 &      0.66 \\
2459250.83505 &      $-$22.03 &      0.50 &       51.59 &      0.72 \\
2459251.76984 &      $-$41.21 &      0.49 &       73.91 &      0.62 \\
2459263.79932 &      $-$60.56 &      0.40 &       93.92 &      0.48 \\
2459264.79986 &       67.57 &      0.54 &      $-$46.83 &      0.64 \\
2459266.76239 &      $-$64.19 &      0.42 &       97.27 &      0.48 \\
2459266.77633 &      $-$64.09 &      0.37 &       96.58 &      0.47 \\
2459272.71189 &      $-$60.26 &      0.39 &       93.33 &      0.50 \\
2459274.72963 &       76.10 &      0.43 &      $-$55.21 &      0.53 \\
2459278.80614 &      $-$57.11 &      0.45 &       90.34 &      0.49 \\
2459329.59451 &       87.87 &      0.39 &      $-$69.73 &      0.44 \\
2459329.73206 &       89.54 &      0.40 &      $-$73.18 &      0.47 \\
%2459330.62106 &      $-$21.83 &       1.05 &       12.66 &       1.43 \\
2459331.62970 &      $-$39.00 &      0.70 &       70.73 &      0.95 \\
2459332.59585 &       84.98 &      0.41 &      $-$68.22 &      0.46 \\

\enddata
\end{deluxetable}

In addition, from this analysis we measured the secondary-to-primary spectroscopic flux ratio in six different orders unaffected by telluric lines between about 770~nm and 890~nm, spanning the center of the TESS bandpass. The average flux ratio obtained is $0.79 \pm 0.05$, which we use below to break the degeneracy between the sum of the stellar radii and the individual radii.

\subsection{Lithium and H\texorpdfstring{$\alpha$}{Ha}}\label{subsec:liha}
As noted above, the spectroscopic observations used to measure the stellar RVs also include the youth indicator \ion{Li}{1} at $\lambda$6708 as well as the activity indicator H$\alpha$ at $\lambda$6563. From the epochs where we were able to easily discriminate the two components without blending, we obtain for H$\alpha$ an equivalent width of $-1.5 \pm 0.1$~\AA\ and $-1.3 \pm 0.1$~\AA\ for the primary and secondary, respectively. For Li we obtain equivalent widths for the primary and secondary, respectively, of $0.30 \pm 0.03$~\AA\ and $0.18 \pm 0.01$~\AA, consistent with the previously reported value of 0.46~\AA\ for the two stars combined \citep{Bowler:2019}. Corrected for dilution by the light of the other star (see Section~\ref{subsec:sed}), we obtain the intrinsic equivalent widths of Li to be $0.53 \pm 0.05$~\AA\ and $0.40 \pm 0.06$~\AA, respectively; for H$\alpha$ we obtain equivalent widths of $-2.6 \pm 0.3$~\AA\ and $-3.0 \pm 0.3$~\AA, respectively. 

In addition, from the widths of the Li lines we obtain a measure of the full-width at half-maximum of $0.78 \pm 0.05$~\AA\ for both components, corresponding to a projected rotational velocity of $v\sin i = 14.7 \pm 1.0$~\kms. As we will see below, the two eclipsing stars have radii of $\approx$0.95~\rsun, such that this $v\sin i$ implies a rotation period of $3.27 \pm 0.22$~d (assuming $i \approx 90^\circ$) for both stars, consistent with synchronous rotation at the orbital period of 3.071~d (see above).

\subsection{Speckle Imaging}\label{subsec:ao}
High-resolution imaging data was acquired for \eb\ with the Zorro speckle instrument on the 8-m Gemini South telescope \citep{Scott:2021} on UT 2022 March 19.
Zorro collects speckle imaging observations simultaneously in two bands (562~nm and 832~nm) with integration times of 60~msec per frame. Seven thousand such observations are obtained and reduced as described in \citet{Howell:2011}, yielding a high-resolution view of the scene near \eb. 

\begin{figure}[!t]
    \centering
    \includegraphics[width=\linewidth,trim=0 0 0 18,clip]{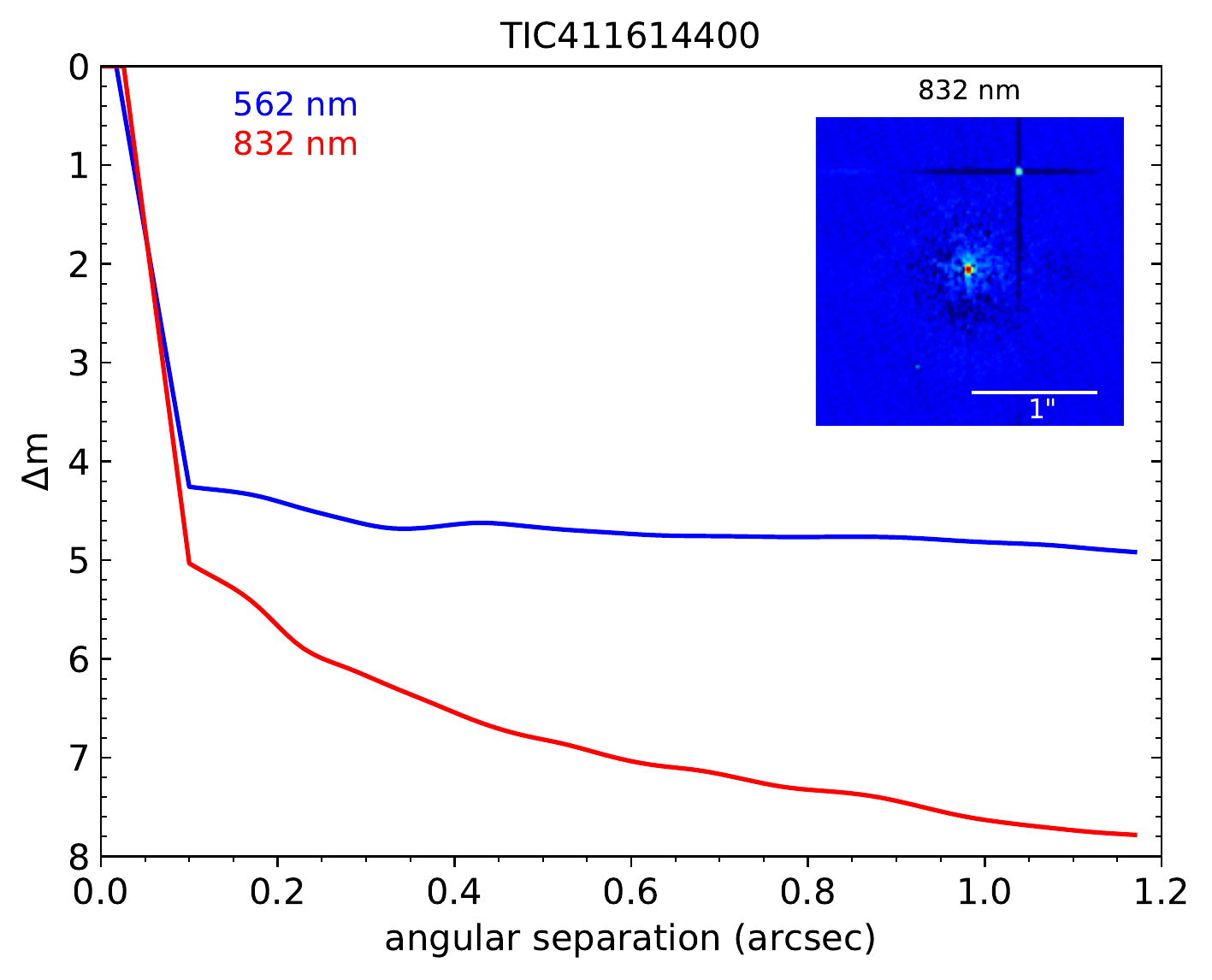}
    \caption{Speckle imaging at 832~nm of \eb\ showing the presence of a faint tertiary companion with $\Delta m = 3.3 \pm 0.3$ at a separation of 0\farcs89 (i.e., $\sim$100~AU physical separation).}
    \label{fig:ao}
\end{figure}

Figure~\ref{fig:ao} shows the resulting 5-$\sigma$ contrast curves obtained in each filter and the reconstructed speckle image in 832~nm. An angularly close stellar companion to \eb\ was discovered in the 832~nm image residing 0\farcs89 away at PA = 333.66$^\circ$. The companion star is $3.3\pm 0.3$ magnitudes fainter than the central EB, and was not reported in any known point source catalogs, including {\it Gaia\/} DR3. 
%\wt{[An extra decimal is needed for the separation error, for consistency. Also in the caption to Fig. 8]}

No other close companions were detected to within a contrast of 5--8 magnitudes for separations ranging from the diffraction limit ($\sim$20~mas) to 1\farcs2.

\section{Analysis and Results}\label{sec:results} 

\subsection{Spectral Energy Distribution: Initial Constraints on Stellar Properties}\label{subsec:sed} 

In order to obtain initial estimates of the component effective temperatures ($T_{\rm eff}$) and radii ($R$), we performed a multi-component fit to the combined-light, broadband SED of the \eb\ system, including broadband photometry from {\it Gaia\/} DR3, 2MASS, and WISE \citep[see][for details of the SED fitting methodology specifically in the context of EBs]{StassunTorres:2016,Miller:2020}.

We started by treating the SED as a single star with $T_{\rm eff}$ estimate based on the spectral type of M1 reported from the spectroscopic observations of \citet{Bowler:2019}. We then redid the fit using two stellar photospheres, their individual $T_{\rm eff}$ and $R$ informed by the eclipse modeling and the total system flux enforced to be consistent with the tight {\it Gaia\/} distance constraint (see Section~\ref{subsec:age}), and then also including the small contribution of light from the faint tertiary companion observed in the speckle imaging (see Section~\ref{subsec:ao}). The resulting $T_{\rm eff}$ and $R$ were iteratively updated based on the joint light-curve and radial-velocity model (Section~\ref{subsec:lcfit}) until a final satisfactory SED fit was produced. 

\begin{figure}[!t]
%    \centering
    \includegraphics[width=0.64\linewidth,trim=130 70 90 90,clip,angle=90]{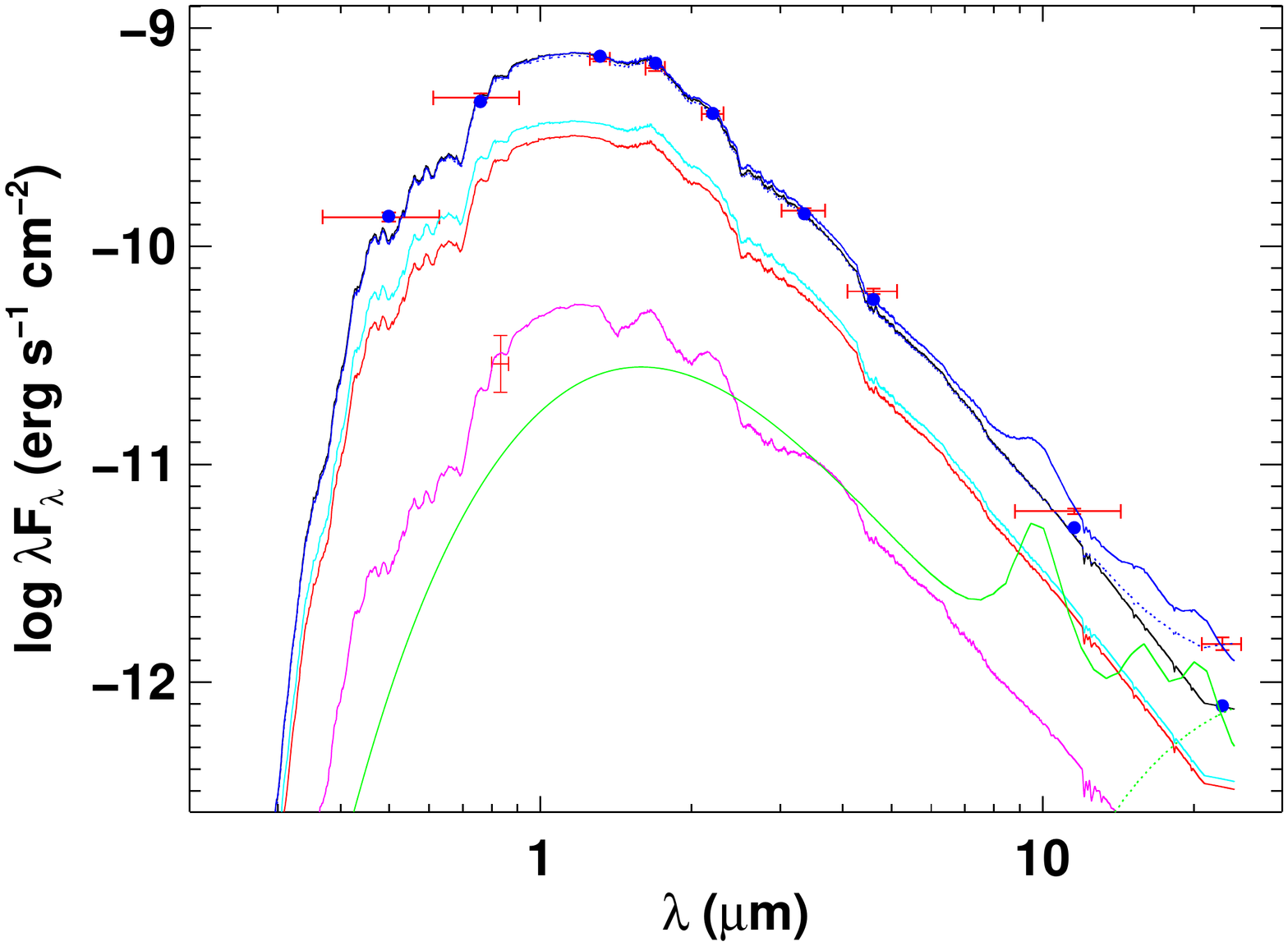}
    \includegraphics[width=\linewidth,trim=52 445 80 70,clip]{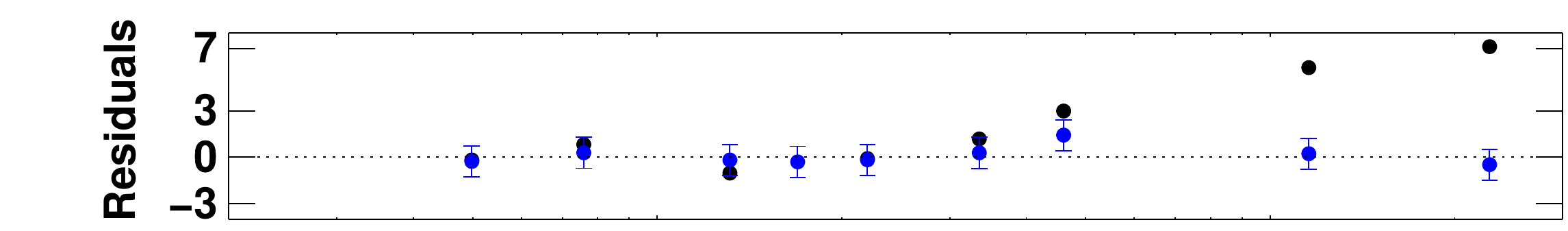}
    \includegraphics[width=0.093\linewidth,trim=70 70 485 90,clip,angle=90]{tic_411614400_sed.pdf}
   \caption{Multi-component fit to the combined-light SED of \eb\ with NextGen stellar atmosphere models. Observed fluxes are represented as red symbols (horizontal error bars represent filter widths); blue symbols are the corresponding model fluxes. The black curve is a single-source fit using  $T_{\rm eff} = 3660$~K, [Fe/H]=0, and $A_V=0$, based on the spectroscopic observations of \citet{Bowler:2019}. Final multi-component fit includes contributions from the primary eclipsing star (cyan curve), the secondary eclipsing star (red curve), the tertiary companion (magenta curve), and the circumbinary disk (green curve). The disk component is represented both without any material within the central cavity (dotted curve) and with $8\times10^{-10}$~\msun\ (i.e., $3\times10^{-4}$~M$_\oplus$) of optically thin dust within the cavity (solid curve). The dark blue curve represents the sum total of all components (dotted and solid for the disk model without and with material in the cavity, respectively). Residuals at bottom are in $\sigma$ units, in the sense of $(O-C)/\sigma_o$ for both the bare stellar model (black) and the full multi-component model (blue).}
    \label{fig:sed}
\end{figure}

As shown in Figure~\ref{fig:sed} (black curve), the SED from 0.4~$\mu$m to 4~$\mu$m can be very well fit by a single component with $T_{\rm eff} = 3660 \pm 100$~K \citep[corresponding to spectral type M1$\pm$1;][]{Bowler:2019}, $\log g \approx 4.3$, [Fe/H]$\,\approx\,$0, and extinction of $A_V \approx 0$ \citep[consistent with other determinations of the reddening to this region; e.g.,][]{Kounkel:2020}. The WISE photometry at 10~$\mu$m and 22~$\mu$m shows what appears to be an excess, which we later model in the context of a circumbinary disk, but which we exclude from the SED fitting at this stage. 

The single-star fit suggests that the two eclipsing components have $T_{\rm eff}$ comparable to one another and close to 3660~K. This is reinforced by the similarity of the primary and secondary eclipses in the TESS light curve, and by the similarity of the strengths of the Li absorption line, though the fact that they are not identical clearly indicates that the two components must have slightly different $T_{\rm eff}$. In fact, the relative eclipse depths imply the ratio of stellar surface brightnesses in the TESS bandpass to be $0.853 \pm 0.002$ (see Section~\ref{subsec:lcfit}), which provides a constraint on the ratio of $T_{\rm eff}$ from the model atmospheres.

An additional constraint is provided by the eclipse durations, which very tightly constrain the sum of the radii of the eclipsing bodies to be $R_{\rm sum} = 1.918 \pm 0.010$~\rsun\ (see Section~\ref{subsec:lcfit}). Another constraint is provided by the ratio of fluxes of the two stars from their relative line strengths in the spectra used to measure the radial velocities (Section~\ref{subsec:rv}), which give $F_2 / F_1 = 0.79 \pm 0.05$ at $\sim$830~nm, averaged over all epochs. Finally, the combined bolometric luminosities of the two stars via the Stefan-Boltzmann relation must reproduce the observed combined-light bolometric flux at Earth given the precise distance provided by {\it Gaia}. 
%Taken together, these constraints remove degeneracies and strongly delimit the permitted stellar $T_{\rm eff}$ and $R$. 

A final iteration was required to self-consistently incorporate the small flux contribution of the tertiary companion (magenta curve in Figure~\ref{fig:sed}). The one direct constraint is provided by the contrast ratio measured in the speckle imaging of the companion relative to the combined light of the EB (Section~\ref{subsec:ao}), represented in Figure~\ref{fig:sed} by the flux measurement at 832~nm fitted by the magenta curve. For the companion we assumed $T_{\rm eff} = 3230$~K and $R = 0.47$~\rsun\ (corresponding to a mass of $\approx$0.18~\msun), guided by the predictions of the \citet{Baraffe:2015} PMS evolutionary models for the nominal 16.2~Myr age of the system (see Section~\ref{subsec:age}). In our best-fit SED model, the tertiary companion contributes 4.5\% of the total light relative to that of the EB, consistent with the estimate of $2.0^{+3.2}_{-1.4}$\% ``third light" ($l_3$) in the eclipse light curve model (Section~\ref{subsec:lcfit}). 

The resulting best-fit SEDs for the two eclipsing stars are represented in Figure~\ref{fig:sed} by the cyan and red curves, respectively, with best-fit $T_{\rm eff}$ of 3749$\pm$35$\pm$11~K and 3645$\pm$35$\pm$11~K, respectively \citep[we adopt an 11~K systematic error for this methodology; see][]{Miller:2020}, best-fit $R$ of 0.981$\pm$0.018~\rsun\ and 0.957$\pm$0.016~\rsun, respectively, and $A_V=0.00^{+0.05}_{-0.00}$. These parameters successfully reproduce the $R_{\rm sum}$ constraint (1.937$\pm$0.019~\rsun\ versus 1.918$\pm$0.010~\rsun\ from the final eclipse model; see Section~\ref{subsec:lcfit}), the spectroscopic flux ratio constraint (0.82$\pm$0.04 versus 0.79$\pm$0.05 from the spectra), the surface brightness ratio constraint (0.851$\pm$0.002 versus 0.853$\pm$0.002 from the average of the 30-min and 2-min cadence light curves in the final eclipse model), and the {\it Gaia\/} distance (105.63$\pm$0.29~pc versus 105.79$\pm$0.28~pc).

The final physical parameters and their precise uncertainty estimates that we ultimately adopt are those provided by the joint MCMC modeling of the eclipse light curves and the radial-velocity measurements described below (Section~\ref{subsec:lcfit}). However, the success of those parameters in reproducing the SED and its various hard constraints is a very strong validation of the global solution for the \eb\ system, which we now discuss.

\subsection{TESS Light Curve Analysis: Determination of Stellar Properties}
\label{subsec:lcfit}

%{\bf WILLIE UPDATE THIS SECTION AND TABLE AS NEEDED.}

We analyzed the TESS light curves of \eb\ using the
Nelson-Davis-Etzel binary model \citep{Etzel:1981, Popper:1981}, as
implemented in the {\tt eb\/} code of \cite{Irwin:2011}. This model is
appropriate for well-detached systems such as this in which the stars
are essentially spherical (see below), and the {\tt eb\/} code
facilitates its use within a Markov chain Monte Carlo (MCMC)
environment. The 30\,m and 2\,m photometry was analyzed jointly with the
radial velocities.

We considered the following adjustable light curve parameters: the
orbital period ($P$), a reference time of primary eclipse ($T_0$), the
central surface brightness ratio in the TESS band ($J \equiv
J_2/J_1$), the sum of the relative radii normalized by the semimajor
axis ($r_1+r_2$), the radius ratio ($k \equiv r_2/r_1$), the cosine of
the orbital inclination angle ($\cos i$), an out-of-eclipse brightness
level in magnitude units ($m_0$), and the eccentricity parameters
$\sqrt{e} \cos\omega_1$ and $\sqrt{e}\sin\omega_1$, where $e$ is the
eccentricity and $\omega_1$ the argument of periastron for the
primary. In view of the noticeable and irregular distortions in the
light curve, we chose to adopt a linear limb-darkening law and allowed
the coefficients ($u_1$, $u_2$) to vary freely; tests with a quadratic law yielded no
improvement, and did not change the geometric quantities. In addition to the
above, the discovery of a close companion to \eb\ described later prompted
us to include an additional parameter, $\ell_3$, to account for third light.
It is defined such that $\ell_1 + \ell_2 + \ell_3 = 1$, in which $\ell_1$
and $\ell_2$ for this normalization are taken to be the light at first
quadrature.
The
spectroscopic parameters in our analysis were the primary and
secondary velocity semiamplitudes ($K_1$ and $K_2$), and the
center-of-mass velocity ($\gamma$).

Because the detrending of the photometry was done independently for
the 30\,m and 2\,m data, as a precaution we allowed separate values of $J$
and $m_0$ to account for possible differences introduced during that process,
as well as the possibility of errors in the contamination corrections.
The normalization of these light curves described in
Section~\ref{subsec:photometry} artificially removes any variations out
of eclipse. Consequently, gravity darkening and reflection become
irrelevant. For consistency, we therefore used the option in the {\tt
eb\/} code that suppresses those effects in calculating the binary
model. Furthermore, as the (flattened) out-of-eclipse portions of the
light curve contain no additional information, we retained only
segments within 0.075 in phase units from the center of each eclipse,
equivalent to about one and a half times the total eclipse duration.

\begin{figure}[!t]
%\epsscale{1.15}
\includegraphics[width=\linewidth,trim=0 0 70 350,clip]{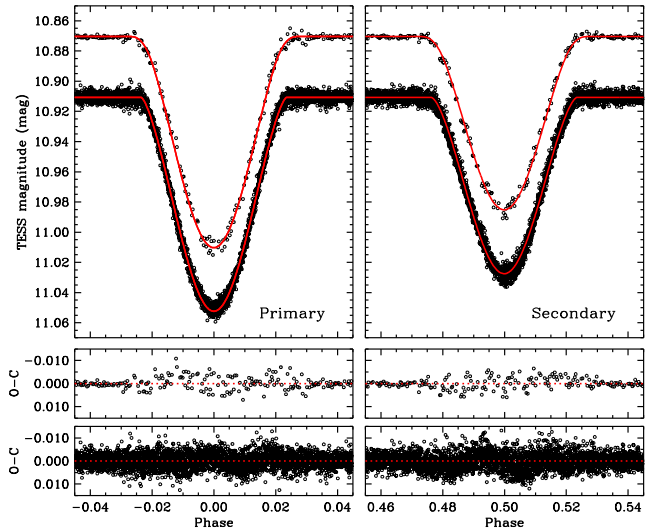}
\figcaption{Detrended TESS observations of \eb\ (2\,m and 30\,m cadence)
  at primary and secondary eclipse. The 30\,m data are displaced
  vertically for clarity, the solid curve is our binary model.
  Residuals for both data sets are shown at bottom.\label{fig:lc}}
\end{figure}

Observational uncertainties were handled by
including four additional free parameters representing multiplicative
scale factors for the internal observational errors in each of the
data sets: the 30\,m and 2\,m photometry, and the primary and secondary
RVs.  The internal errors for the RVs are those given in
Table~\ref{tab:rvs}, and for the 30\,m and 2\,m photometry we adopted
values of 0.002 and 0.003~mag, respectively.  These four additional
parameters were solved simultaneously and self-consistently with the
other variables \citep[see, e.g.,][]{Gregory:2005}.

We carried out our joint lightcurve and RV analysis using the {\tt emcee\/}
%\footnote{\url{https://github.com/dfm/emcee}} 
code of
\cite{Foreman-Mackey:2013}, which is a Python implementation of the
affine-invariant MCMC ensemble sampler proposed by
\cite{Goodman:2010}.  We used 100 walkers with chain lengths of 20,000
each, after discarding the burn-in. All parameters used uniform or
log-uniform priors over suitable ranges that are listed in the last column of
Table~\ref{tab:mcmc}. We verified convergence by visual examination of
the chains, and requiring a Gelman-Rubin statistic of 1.05 or
smaller for each parameter \citep{Gelman:1992}. The cadence of the
TESS FFIs (30\,m) corresponds to a fraction of the 3.07~day orbital period
of \eb\ that is not quite negligible, equivalent to about
0.007 in phase units. To avoid biases from smearing, we oversampled
the model light curve at each iteration of our solution, and then
integrated over the 30\,m duration of each cadence prior to the
comparison with the observations \citep[see][]{Gilliland:2010,
  Kipping:2010}.

\begin{figure}[!t]
%\epsscale{1.15}
\includegraphics[width=\linewidth,trim=0 0 70 240,clip]{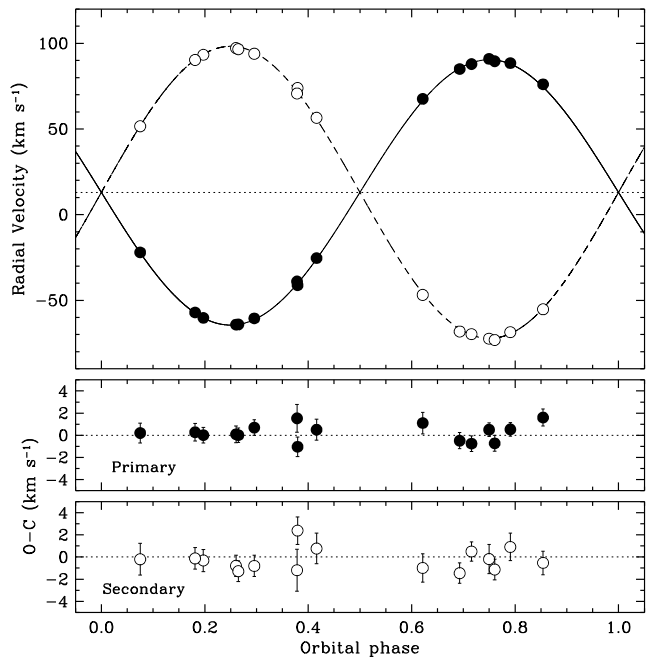}
\figcaption{Radial-velocity measurements of \eb\ with our adopted
  model. Primary and secondary observations are represented with
  filled and open circles, respectively, the dotted line marks the
  center-of-mass velocity of the system. Error bars are smaller than
  the symbol sizes. Residuals are shown at bottom.\label{fig:rvs}}
\end{figure}

Initial solutions revealed that the radius ratio $k$ was poorly
constrained by the photometry. This is often the case in eclipsing
binaries such as \eb\ with partial eclipses and similar components,
and is caused by strong correlations with other free parameters.  For
example, in our case the correlation coefficient between $k$ and $\cos
i$ was $-0.989$, and between $k$ and $r_1+r_2$ it was $-0.984$.  A
common remedy is to require the lightcurve solution to be consistent
with an independently measured light ratio, such as from spectroscopy
\citep[see, e.g.,][]{Andersen:1980}. This is effective because the
light ratio depends very strongly on the radius ratio: $\ell_2/\ell_1
\propto k^2$. We therefore applied our flux ratio from CHIRON ($0.79
\pm 0.05$, measured at wavelengths near the center of the TESS
bandpass) as a Gaussian prior, and this largely removed the
degeneracy.

It was also noticed that the residuals of the primary velocities were
predominantly positive, and those of the secondary mostly negative. We
speculate that this unphysical effect may be a consequence of
distortions in the spectral line profiles produced by spots, which can
then affect the velocities. For our final solution we chose to solve
for separate values of $\gamma$ for each component, which removed the
bias.

\setlength{\tabcolsep}{3pt}
\begin{deluxetable}{lcc}
\tablewidth{0pt}
\tablecaption{Joint Photometric-Spectroscopic Solution for \eb}\label{tab:mcmc}
\tablehead{ \colhead{~~~~~~~Parameter~~~~~~~} & \colhead{Value} & \colhead{Prior} }
\startdata
 $P$     (days)               &  $3.07165752^{+0.00000034}_{-0.00000033}$   & [3.0, 3.1] \\ [1ex]
 $T_0$\tablenotemark{a}       &  $571.768221^{+0.000080}_{-0.000081}$     & [571.7, 572.0] \\ [1ex]
 $J$, 30\,m data              &  $0.951^{+0.039}_{-0.017}$                  & [0.4, 1.2] \\ [1ex]
 $J$, 2\,m data               &  $0.955^{+0.039}_{-0.017}$                  & [0.4, 1.2] \\ [1ex]
 $r_1+r_2$                    &  $0.19295^{+0.00078}_{-0.00167}$            & [0.01, 0.40] \\ [1ex]
 $k \equiv r_2/r_1$           &  $0.966^{+0.021}_{-0.023}$                  & [0.5, 2.0] \\ [1ex]
 $\cos i$                     &  $0.1258^{+0.0011}_{-0.0025}$               & [0, 1] \\ [1ex]
 $\sqrt{e} \cos\omega_1$      &  $-0.0049^{+0.0023}_{-0.0063}$              & [$-$1, 1] \\ [1ex]
 $\sqrt{e} \sin\omega_1$      &  $+0.033^{+0.037}_{-0.045}$                 & [$-$1, 1] \\ [1ex]
 $u_1$                        &  $0.121^{+0.051}_{-0.057}$                  & [0.0, 1.0] \\ [1ex]
 $u_2$                        &  $0.429^{+0.037}_{-0.040}$                  & [0.0, 1.0] \\ [1ex]
 $m_{\rm 0,\,30min}$ (mag)      &  $10.870415^{+0.000083}_{-0.000082}$        & [10, 11] \\ [1ex]
 $m_{\rm 0,\,2min}$ (mag)       &  $10.870696^{+0.000035}_{-0.000034}$        & [10, 11] \\ [1ex]
 $\ell_3$                     &  $0.020^{+0.032}_{-0.014}$                  & [0.0, 0.5] \\ [1ex]
 $\gamma_1$ (\kms)            &  $+13.18^{+0.19}_{-0.19}$                   & [0, 20 ]  \\ [1ex]
 $\gamma_2$ (\kms)            &  $+12.63^{+0.27}_{-0.27}$                   & [0, 20 ]  \\ [1ex]
 $K_1$ (\kms)                 &  $77.44^{+0.21}_{-0.21}$                    & [50, 100 ]   \\ [1ex]
 $K_2$ (\kms)                 &  $85.26^{+0.30}_{-0.30}$                    & [50, 100 ]   \\ [1ex]
 $f_{\rm 30\,m}$              &  $1.005^{+0.030}_{-0.028}$                  & [$-$5, 5] \\ [1ex]
 $f_{\rm 2\,m}$               &  $0.9944^{+0.0069}_{-0.0069}$               & [$-$5, 5] \\ [1ex]
 $f_{\rm RV1}$                &  $1.73^{+0.45}_{-0.27}$                     & [$-$5, 5] \\ [1ex]
 $f_{\rm RV2}$                &  $1.89^{+0.51}_{-0.30}$                     & [$-$5, 5] \\ [1ex]
\noalign{\hrule} \\ [-3ex]
\multicolumn{3}{c}{Derived quantities} \\ [0.25ex]
\noalign{\hrule} \\ [-2.5ex]
 $r_1$                        &  $0.0980^{+0.0012}_{-0.0012}$               & \nodata \\ [1ex]
 $r_2$                        &  $0.0946^{+0.0012}_{-0.0015}$               & \nodata \\ [1ex]
 $i$ (degree)                 &  $82.773^{+0.146}_{-0.061}$                 & \nodata \\ [1ex]
 $e$                          &  $0.0012^{+0.0037}_{-0.0010}$               & \nodata \\ [1ex]
 $\ell_2/\ell_1$, 30\,m data  &  $0.798^{+0.035}_{-0.035}$                  & $G(0.79, 0.05)$ \\ [1ex]
 $\ell_2/\ell_1$, 2\,m data   &  $0.801^{+0.035}_{-0.035}$                  & $G(0.79, 0.05)$ \\ [1ex]
 $J_{\rm ave}$, 30\,m data    &  $0.852^{+0.021}_{-0.008}$                  & \nodata \\ [1ex]
 $J_{\rm ave}$, 2\,m data     &  $0.855^{+0.021}_{-0.006}$                  & \nodata \\ [1ex]
\enddata
\tablecomments{Values correspond to the mode of the
  posterior distributions, uncertainties represent 68.3\%
  credible intervals. Priors in square brackets are uniform over the
  specified ranges, except the ones for $f_{\rm 30\,m}$, $f_{\rm 2\,m}$,
  $f_{\rm RV1}$ and $f_{\rm RV1}$, which are log-uniform, and the ones
  for the light ratios, which are Gaussian, indicated as $G({\rm
    mean}, \sigma)$.}
\tablenotetext{a}{Time units are BJD$-$2,458,000.}
\end{deluxetable}
\setlength{\tabcolsep}{6pt}

Table~\ref{tab:mcmc} presents the results of the analysis. We list the
mode of the posterior distribution for each parameter, along with the
corresponding 68.3\% credible intervals. The posterior distributions
for the derived quantities in the bottom section of the table were
constructed directly from the MCMC chains of the adjustable parameters
involved. The eccentricity of the orbit is not statistically
significant. We derive a 3$\sigma$ upper limit of $e \approx 0.01$.
The oblateness indices for the stars, calculated following the
prescription by \cite{Binnendijk:1960}, are 0.0013 for the primary and
0.0014 for the secondary. These are well below the upper limit of 0.04
considered safe for the Nelson-Davis-Etzel binary model \citep[see,
  e.g.,][]{Popper:1981}, justifying its use.  The photometric
observations are shown together with our model in
Figure~\ref{fig:lc}. The corresponding graphic for the radial
velocities is presented in Figure~\ref{fig:rvs}.

Our mass and radius determinations for \eb\ are among the best
determined for PMS stars. The masses have relative precisions of 0.8 and
0.7\% for the primary and secondary, and the absolute radii are
determined to 1.3\% and 1.6\%, respectively. We list these results in
Table~\ref{tab:dimensions} along with other derived properties. The
bolometric luminosities listed in the table are from the Stefan-Boltzmann relation, based on these radii together with the
effective temperatures as determined from the SED fit (see Section~\ref{subsec:sed}).

As a sanity check, the distance to the system may be calculated directly
(and independently from {\it Gaia}) from the luminosities, the
apparent visual magnitude \citep[$V = 11.468 \pm 0.087$;][]{Kiraga:2012}, and bolometric
corrections $BC_{\rm V}$ of $-1.30$ and $-1.44$ for the primary and
secondary, respectively, based on the work of \citet{Eker:2020}. The extinction
is negligible according to our SED fit ($A_{\rm V} < 0.05$).

The value we obtain, $104.7 \pm 9.8$~pc, corresponds to a parallax of $9.49 \pm 0.89$~mas which,
although much less precise than the determination from {\it Gaia\/} DR3 ($9.453 \pm
0.025$~mas; see Table~\ref{tab:dimensions}) is in good agreement with it. The large
error in the distance in this calculation is driven by the errors in $V$ (affected
by the intrinsic variability), the bolometric corrections, and the adopted uncertainty
in the interstellar extinction.

Finally, on the assumption of spin-orbit synchronization and alignment, the
predicted rotational velocities are near 16~\kms\ for both stars. This
is fairly close to our spectroscopic determination 
%based on the Li lines
(Section~\ref{subsec:rv}).

\subsection{Photometric Variability}\label{subsec:var}

As noted in Section~\ref{subsec:photometry}, the TESS light curve of \eb\ exhibits variability beyond the primary and secondary eclipses, at a level of $\sim$5\% peak to peak. 
%\wt{[I don't see mention of this in Section 3.1]}. 
We attempted to characterize these variations in the context of rotationally modulated spot variations on one or more of the stars in the system, as such spot-driven variations at levels of a few percent to nearly 50\% peak-to-peak are frequently observed among low-mass PMS stars \citep[see, e.g., ][for examples]{Stassun:1999}.

\setlength{\tabcolsep}{2pt}
\begin{deluxetable}{lcc}
%\tablewidth{0pc}
%\tablewidth{1.0\columnwidth}
\tablecaption{Physical Properties of \eb \label{tab:dimensions}}
\tablehead{ \colhead{~~~~~~~~~~Parameter~~~~~~~~~~} & \colhead{Primary} & \colhead{Secondary} }
\startdata
 $M$ ($\mathcal{M}_{\sun}^{\rm N}$)         &  $0.7354 \pm 0.0057$           &  $0.6680 \pm 0.0044$  \\ 
 $R$ ($\mathcal{R}_{\sun}^{\rm N}$)         &  $0.976 \pm 0.013$             &  $0.942 \pm 0.015$    \\ 
 $\log g$ (dex)                             &  $4.33 \pm 0.011$              &  $4.31 \pm 0.013$     \\ 
 $q \equiv M_2/M_1$                         &          \multicolumn{2}{c}{$0.9083 \pm 0.0041$}       \\
 $a$ ($\mathcal{R}_{\sun}^{\rm N}$)         &          \multicolumn{2}{c}{$9.956 \pm 0.023$}         \\ 
 $T_{\rm eff}$ (K)                          &  $3749 \pm 35$                 &  $3645 \pm 35$        \\ 
 $L$ ($L_{\sun}$)                           &  $0.169 \pm 0.008$             &  $0.141 \pm 0.007$    \\ 
 $M_{\rm bol}$ (mag)                        &  $6.67 \pm 0.05$               &  $6.87 \pm 0.05$      \\ 
 $BC_V$ (mag)                               &  $-1.30 \pm 0.21$              &  $-1.44 \pm 0.21$     \\ 
 $M_V$ (mag)                                &  $7.97 \pm 0.22$               &  $8.31 \pm 0.22$      \\ 
 $v_{\rm sync} \sin i$ (\kms)\tablenotemark{a} &  $15.94 \pm 0.21$           &  $15.40 \pm 0.24$     \\ 
 $v \sin i$ (\kms)\tablenotemark{b}         &  $14.7 \pm 1.0$                &  $14.7 \pm 1.0$       \\ 
 $E(B-V)$ (mag)                             &          \multicolumn{2}{c}{$< 0.02$}                  \\ 
 $A_V$ (mag)                                &          \multicolumn{2}{c}{$< 0.05$}                  \\ 
% Distance modulus (mag)                     &          \multicolumn{2}{c}{$5.11 \pm 0.19$}           \\ 
% Distance (pc)                              &          \multicolumn{2}{c}{$104.7 \pm 9.8$}              \\ 
% $\pi$ (mas)                                &          \multicolumn{2}{c}{$9.49 \pm 0.89$}           \\ 
 $\pi_{Gaia/{\rm DR3}}$ (mas)\tablenotemark{c} &      \multicolumn{2}{c}{$9.453 \pm 0.025$}          \\
 $d_{Gaia/{\rm DR3}}$ (pc)\tablenotemark{c} &      \multicolumn{2}{c}{$105.79 \pm 0.28$}          \\ [0.5ex]
\enddata
\tablecomments{The masses, radii, and semimajor axis $a$ are expressed
  in units of the nominal solar mass and radius
  ($\mathcal{M}_{\sun}^{\rm N}$, $\mathcal{R}_{\sun}^{\rm N}$) as
  recommended by 2015 IAU Resolution B3 \citep[see][]{Prsa:2016}, and
  the adopted solar temperature is 5772~K (2015 IAU Resolution B2).
  Bolometric corrections are from the work of \citet{Eker:2020}. See text for the
  source of the reddening. For the apparent visual magnitude of
  \eb\ out of eclipse we used $V = 11.468 \pm 0.087$
  \citep{Kiraga:2012}. }
\tablenotetext{a}{Synchronous projected rotational velocity assuming a
  circular orbit and spin-orbit alignment.}
\tablenotetext{b}{Measured projected rotational velocities.}
\tablenotetext{c}{A parallax zero-point correction of $+0.004$~mas has
  been added to the parallax \citep{Lindegren:2021} and a scaling
 factor of 1.20 has been applied to the internal error, following
\citet{ElBadry:2021}.}
\end{deluxetable}
\setlength{\tabcolsep}{6pt}

\begin{figure*}[!t]
    \centering
    \includegraphics[width=\linewidth]{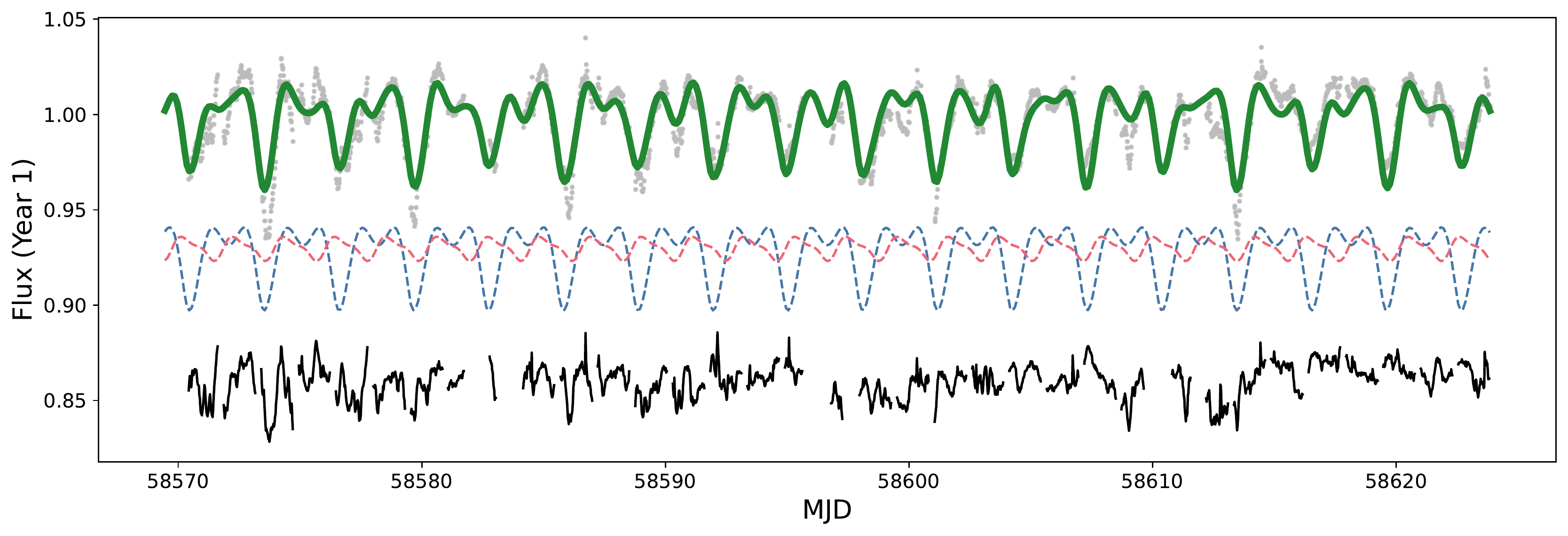}
    \includegraphics[width=\linewidth]{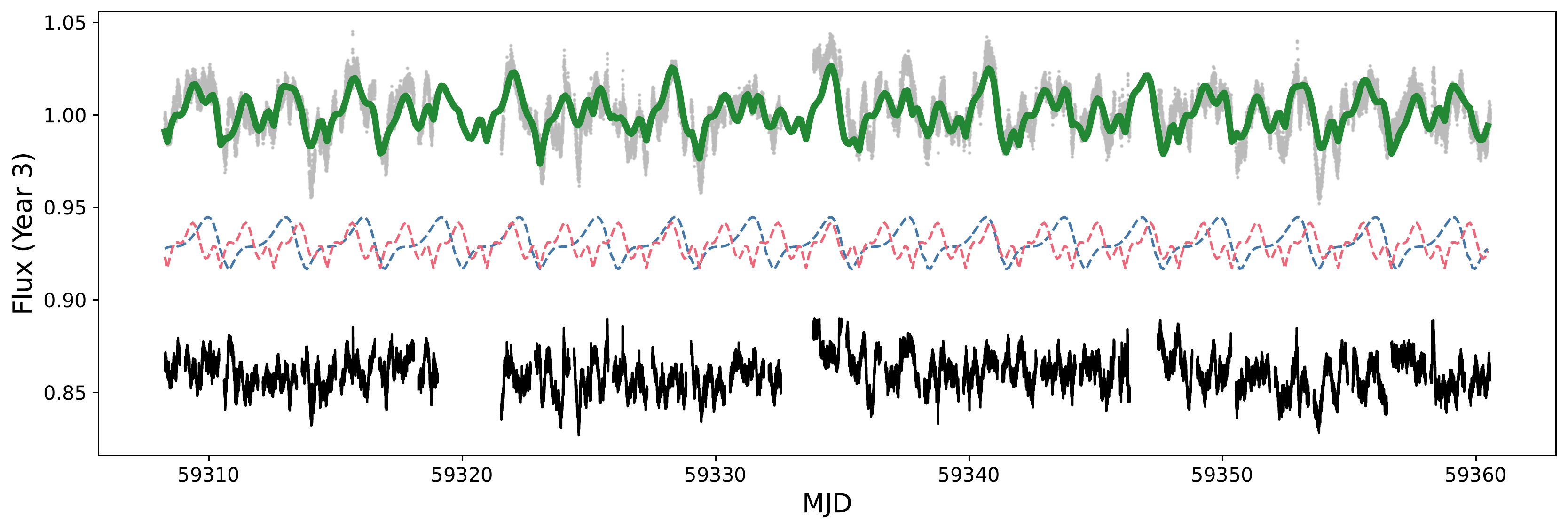}
    \caption{Out-of-eclipse variations in the TESS light curve of \eb\ for Sectors 10--11 (top) and Sectors 37--38 (bottom). While there is some change in overall morphology of the variations, at both epochs the variations can be characterized with two dominant periodic components: a ``double-humped" component at the orbital period that we attribute to rotationally modulated spot variations on one or both of the eclipsing stars (blue dashed curve) and a somewhat more complex, sawtooth-like component with a period of 2.1~d that we attribute to rotationally modulated spot variations on the tertiary companion (red dashed curve). Both periodic components added together are shown in green overlaid on the TESS light curve data (grey symbols). The residuals (black, bottom) reveal additional, quasi-stochastic variations at the level of $\sim$2\% peak-to-peak that may be attributable to accretion in the system from the circumbinary disk (see Section~\ref{sec:discussion}).}
    \label{fig:ooe_var}
\end{figure*}

To model this variability, we first removed all the fluxes corresponding to the eclipses. We then folded the light curve to the dominant period found with a Lomb-Scargle periodogram, which was the same as the orbital period, or $\sim$3.07~d. We then fitted an 8th degree polynomial to this folded light curve and subtracted it out. This was done independently to the Year~1 data corresponding to Sectors 10--11, and Year~3 data corresponding to Sectors 37--38. 

%\vfill\eject 

Significant variability remained in the residuals, showing a secondary periodic signature, with a period of $\sim$2.1~d, consistent both in Year~1 and Year~3. It was similarly fit with an 8th degree polynomial in the newly folded residual light curve, independently in Year~1 and Year~3 data. The resulting fit is shown in Figure~\ref{fig:ooe_var}.

There was some change in overall morphology of the variations during the 2 years that elapsed between TESS Sectors 10--11 and Sectors 37--38, with Year~1 data having a larger amplitude of variability by almost a factor of 2 compared to Year~3 data. Additionally, the Year~1 data show a ``simpler'' morphology, with fewer harmonic oscillations compared to Year~3. Nonetheless, in both time ranges, the primary period component appears to be ``double-humped", and the secondary component at a period of 2.1~d shows a sawtooth-like morphology. 

Adding the two periodic signatures does not fully model the variability. The residuals reveal additional, quasi-stochastic ``dip-like" variations at the level of $\sim$2\% peak-to-peak. While they are aperiodic, 
%their typical duration roughly corresponds to 1/3 of the secondary period, or roughly 0.7~d.
they appear to have a typical duration of $\sim$0.3~d. 

We may attribute the periodic components of the variability to rotational modulation of spots on the stars in the system. That one of the periodic components shares the orbital period of the eclipsing stars strongly suggests that it can be ascribed to one or both of the eclipsing stars, whose rotational periods are synchronized with the orbit. The fact that this component appears double-humped could mean either that this signal arises from spots on roughly opposite hemispheres on one of the stars, or that it arises from spots on both stars at roughly opposite longitudes. The observed peak-to-peak amplitude of $\sim$5\% implies that the intrinsic amplitude is either $\sim$5\% for both stars or $\sim$10\% if it arises from one star and diluted by the light of the other. 

The second periodic signal cannot readily be attributed to the eclipsing stars. The observed period of 2.1~d is not an obvious harmonic or alias of the orbital period. The more likely source is rotationally modulated spot variations on the tertiary companion. In that case, the observed amplitude of $\sim$1\% implies an intrinsic amplitude of $\sim$20\%, given the tertiary's light is diluted by a factor of $\sim$20 by the eclipsing stars. 

Finally, the residual quasi-stochastic variations may be attributable to activity or accretion in the system. We defer discussion of this possibility to Section~\ref{subsec:accretion}. 

\section{Discussion}\label{sec:discussion} 

\subsection{Comparison with Stellar Evolution Models}\label{subsec:models}

\begin{figure}[!t]
%\epsscale{1.15}
\includegraphics[width=\linewidth,trim=0 0 95 180,clip]{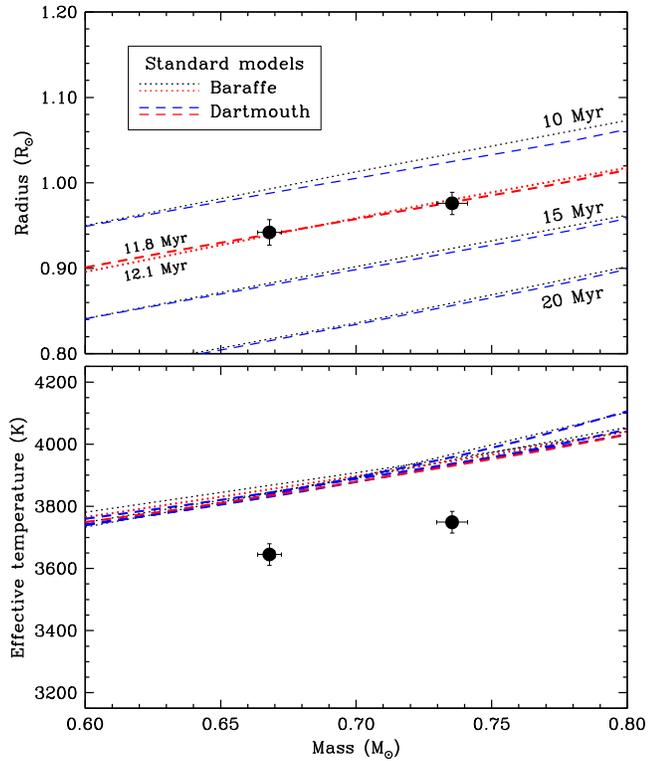}
\figcaption{Mass, radius, and effective temperature determinations for the eclipsing binary components of \eb\ compared
against standard stellar evolution models by \cite{Baraffe:2015} and the Dartmouth series \citep{Dotter:2008, Feiden:2016}. Isochrone ages are labeled. The best match to the radii (shown in red) is obtained at about 12~Myr for both models, but theoretical predictions appear too hot at this age.\label{fig:std_models}}
\end{figure}

Our precise mass, radius, and temperature determinations offer an opportunity to compare the measurements against current models of stellar evolution for young objects. Figure~\ref{fig:std_models} shows our determinations for the primary and secondary against standard model isochrones from \cite{Baraffe:2015} and from the Dartmouth series \citep{Dotter:2008, Feiden:2016}. The best fit age based on our measured radii and masses is near 12~Myr for both sets of models (top panel). However, the effective temperatures appear too cool at these masses (bottom panel), deviating from theoretical predictions by about 5$\sigma$.

As mentioned earlier, stellar activity has been found to affect the global properties of stars with convective envelopes \citep[see, e.g., the reviews by][]{Torres:2013, Feiden:2015}, often causing them to have larger radii and cooler temperatures than models indicate. One explanation invokes magnetic fields, which are commonly associated with activity and inhibit the convective flow of energy, to which stars respond by increasing their surface area and lowering their temperature. Non-standard models that incorporate magnetic fields have been successful in explaining radius inflation and temperature suppression in several eclipsing main-sequence systems with well measured properties \citep[see, e.g.,][]{Feiden:2013, MacDonald:2014}, and similar models have also been explored for the far fewer eclipsing PMS binaries that have sufficiently precise determinations of their masses, radii, and temperatures \citep[see, e.g.,][]{Stassun:2014}.

\begin{figure}[!t]
%\epsscale{1.15}
\includegraphics[width=\linewidth,trim=0 0 95 180,clip]{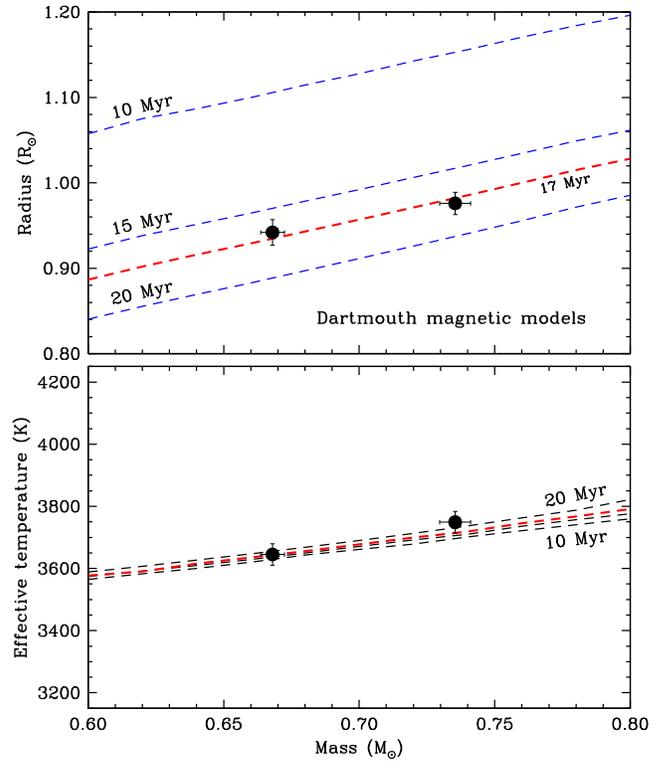}
\figcaption{Same as Figure~\ref{fig:std_models} (and on the same scale), now comparing the measured properties of \eb\ against a version of the Dartmouth models that incorporates the effects of magnetic fields \citep{Feiden:2016}. Both the radii and temperatures of the two components are well matched by theory at an age of 17~Myr.\label{fig:magnetic_models}}
\end{figure}

Figure~\ref{fig:magnetic_models} compares the measured properties of \eb\ against magnetic models by
\cite{Feiden:2016}, which are an evolution of the same Dartmouth models shown in Figure~\ref{fig:std_models} in which the internal structure equations have been modified to account for magnetic pressure. In this case the age that matches the measured radii of \eb\ is somewhat older than before ($17.0 \pm 0.5$~Myr), but in better agreement with the age estimated for the LCC association (see Section~\ref{subsec:age}), and now the same isochrone also reproduces the temperatures of both components within the uncertainties. 
%This example provides valuable confirmation that magnetic models are able to account for the effects of stellar activity at this young age.

\subsection{Activity and Accretion}\label{subsec:accretion}
\begin{figure}[!t]
    \centering
    \includegraphics[width=\linewidth]{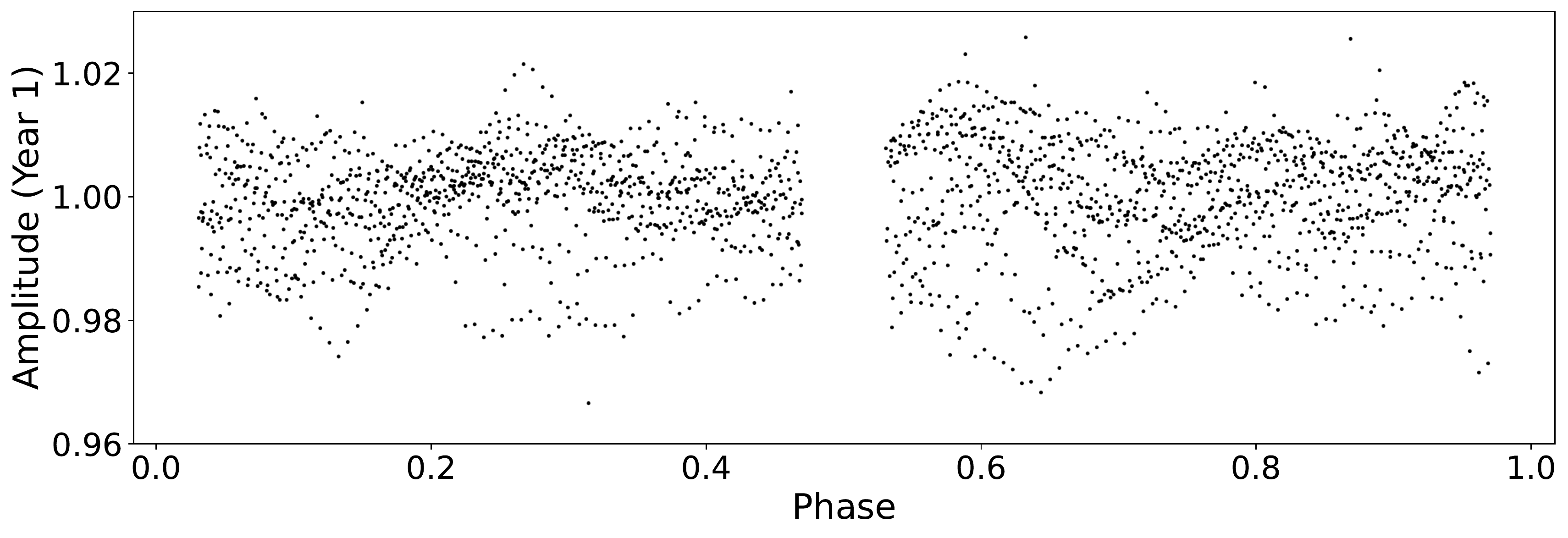}
    \includegraphics[width=\linewidth]{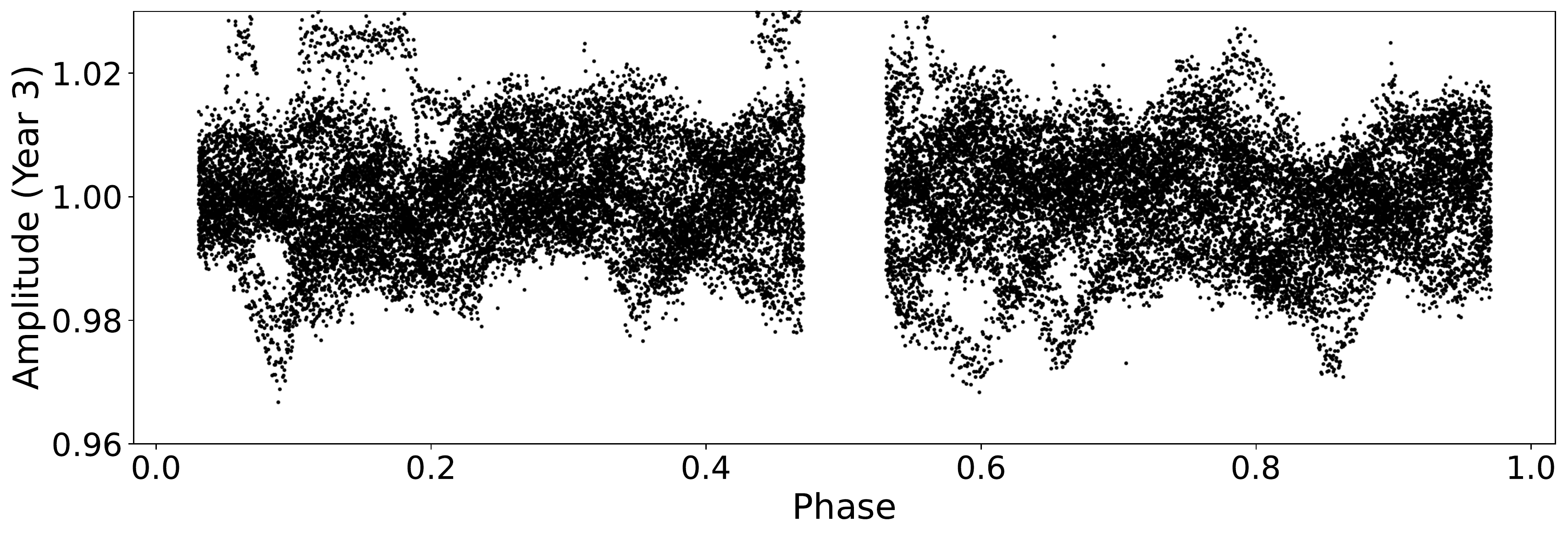}
    \caption{Variability residuals in the TESS light curve data in Sectors 10--11 (top) and Sectors 37--38 (bottom).}
    \label{fig:lcresid}
\end{figure}

The TESS light curve of \eb\ exhibits quasi-stochastic photometric variability beyond the eclipses and beyond the additional periodic variations that we attribute to spots on the eclipsing stars and the tertiary companion (see Section~\ref{subsec:var}). A detailed view of these variations as a function of orbital phase is shown in Figure~\ref{fig:lcresid}. 
The variations are complex and exhibit changes in morphology on short timescales. However, there is a recurring ``dip" type behavior reminiscent of the persistent and transient flux dips that have now been observed in a large number of young stars \citep[see, e.g.,][]{Rebull:2015,Stauffer:2014,Stauffer:2015,Stauffer:2017,Stauffer:2021}, of the type which has been attributed in previous studies to transits of the stars by clumps in their protoplanetary disks and/or material in accretion streams. 
%There is a hint in Figure~\ref{fig:lcresid} that the more prominent dips tend to occur soon after the eclipses, especially secondary eclipse, although this is difficult to quantify. There may also be a tendency for brief brightenings or flares to occur on either side of secondary eclipse.

\begin{figure}[!t]
    \centering
    \includegraphics[width=\linewidth,trim=70 548 60 80,clip]{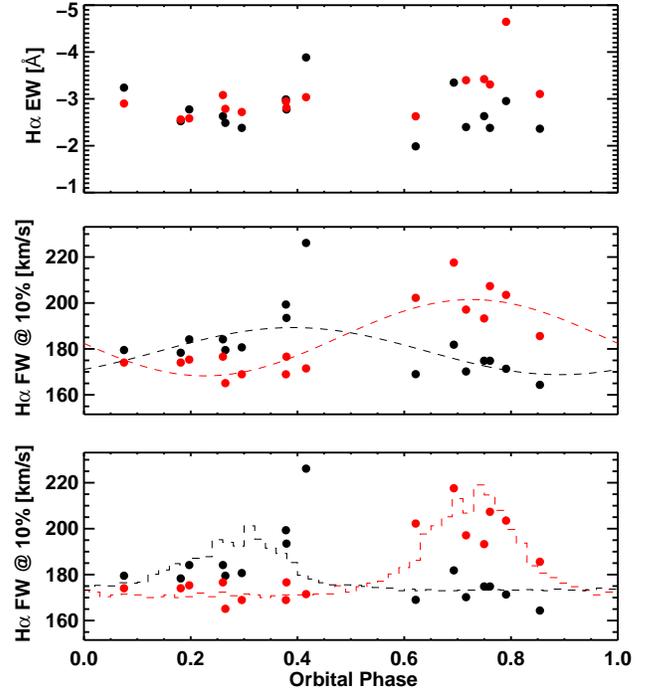} 
    \includegraphics[width=\linewidth,trim=70 335 60 295,clip]{ha_var_new.pdf}
    \includegraphics[width=\linewidth,trim=70 80 60 505,clip]{ha_var_new.pdf}
    \caption{Strength of H$\alpha$ emission as a function of orbital phase for the primary (black symbols) and secondary (red symbols) eclipsing stars. Equivalent width (EW) variations (top) do not exhibit periodic or other clear behavior with orbital phase. Full-width (FW) at 10\% intensity variations (middle) of the secondary do exhibit statistically significant periodic behavior, represented by a best-fit sinusoid (red dashed curve); a best-fit sinusoid for the primary (black dashed curve) is also shown for comparison. (Bottom) Same as middle, except curves represent theoretical predictions from the simulations of \citet{Artymowicz:1996} for the relative accretion rate of circumbinary disk material onto the two stars. The vertical scale for the model curves is arbitrary but the curves have been normalized to preserve the relative accretion rate changes in the model. The black curve representing the model accretion variations for the primary has been shifted by 0.5 phase relative to that for the secondary (red) to simulate the effect of seeing only one longitudinal hemisphere (or ``side") of each star at each orbital quadrature; see the text.}
    \label{fig:havar}
\end{figure}

We also examined the variability of the H$\alpha$ emission lines present in the spectra we used to measure the radial velocities (Section~\ref{subsec:rv}). Using the spectra from epochs when the two eclipsing stars are well separated in radial velocity, we measured the H$\alpha$ equivalent widths as well as the the full-width at 10\% intensity. The former is a commonly used diagnostic for chromospheric activity, the latter has been demonstrated to be a more reliable diagnostic for accretion-related activity specifically \citep[see, e.g.,][]{White:2003}. Both are shown in Figure~\ref{fig:havar} as a function of orbital phase.

The H$\alpha$ equivalent width variations do not show any clear behavior with orbital phase, and a Lomb-Scargle period search finds no significant periodicity. The H$\alpha$ equivalent widths may therefore be principally a manifestation of the stellar chromospheres which, while exhibiting some degree of true variability, are evidently not sufficiently well organized to produce clear rotationally modulated variations. We return to the implications of the chromospheric activity in Section~\ref{subsec:chromo}. 

The H$\alpha$ full widths at 10\% intensity, on the other hand, do show evidence for variations on the orbital period. A Lomb-Scargle period search applied to the secondary's variations independently finds a best period of 3.07~d (i.e., identical to the orbital period) with a false-alarm probability of 0.03. A best-fit sinusoid is represented in Figure~\ref{fig:havar} to guide the eye, but is not intended to suggest that the variations are in fact sinusoidal. More generically, a Student's $t$ test applied to the secondary's measurements on either side of orbital phase 0.5 finds that the means are significantly different with 99.98\% confidence.

A Lomb-Scargle search applied to the primary's variations in full-width at 10\% intensity does not identify any statistically significant periodicity. The primary's variations do show one measurement near orbital phase 0.4 (i.e., just prior to secondary eclipse) that could be due to a particularly strong accretion event. In addition, a Student's $t$ test finds, as with the secondary, that the primary's measurements on either side of orbital phase 0.5 have significantly different means with 99\% confidence. Therefore, we again show a best-fit sinusoid on the orbital period in Figure~\ref{fig:havar} to guide the eye. 

It is interesting that the full-width at 10\% intensity variations, assuming they are accretion driven, are aware of the orbital period. Moreover, the times of relatively stronger accretion on the primary and secondary appear to occur near the quadrature phases, when the stars are side by side as seen by the observer, and nearly half an orbital period apart: The best-fit sinusoids in Figure~\ref{fig:havar} for the primary and secondary peak at phases of $\sim$0.35 and $\sim$0.75, respectively (i.e., separated by $\sim$0.4 in phase). These behaviors are consistent with what has been seen in a few other young binaries that exhibit orbitally pulsed accretion from a circumbinary disk via accretion streams that reach the stars across the hole in the disk carved out by the binary \citep[e.g.,][]{Mathieu:1997,Jensen:2007,Ardila:2015}. 

The canonical examples are DQ~Tau \citep{Mathieu:1997,Tofflemire:2017dq,Muzerolle:2019}, UZ~Tau~E \citep{Jensen:2007}, and TWA~3A \citep{Tofflemire:2017twa,Tofflemire:2019}. In all cases, the central binary is surrounded by a circumbinary disk whose SED is best modeled by a passive dusty disk with an inner hole that is comparable in size to the binary orbit and that includes a small amount of warm, optically thin dust within the hole. In all cases, photometric variations as well as variations in H$\alpha$ and other spectroscopic accretion indicators \citep[see][]{Ardila:2015} vary periodically with the orbital period in a manner consistent with theoretical predictions \citep[see, e.g.,][]{Artymowicz:1996}. 

In Figure~\ref{fig:havar} we represent the same model predictions from \citet{Artymowicz:1996} for the accretion rate variations as a function of orbital phase that were used by \citet{Mathieu:1997} for comparison with their observations of DQ~Tau. We chose that model because it is the case studied by \citet{Artymowicz:1996} that has a binary with nearly equal-mass stars, as is the case for the system under consideration here ($q \approx 0.9$; see Table~\ref{tab:dimensions}). Note, however, that the DQ~Tau binary has a considerably longer orbit than \eb\ (15.8~d versus 3.07~d) and its orbit is also eccentric. Perhaps most importantly, the DQ~Tau binary does not eclipse; in fact, it is likely that it is viewed at a nearly pole-on inclination angle \citep[][]{Basri:1997,Czekala:2016}, such that variability arising from accretion stream footpoints on the stellar surfaces may be more easily viewed throughout the orbit. In \eb, only one longitudinal hemisphere (or ``side") of each star can be viewed at each orbital quadrature. 

We applied an offset of 0.25 phase to the model accretion rate curves in Figure~\ref{fig:havar} to match the apparent quadrature phasing of the data. In the model, however, the highest accretion rates onto the two stars occur at the same orbital phase, whereas in \eb\ the data imply that the highest accretion rates observed for the primary occur nearly 0.5 phase apart from those observed for the secondary. Therefore, we applied an additional offset of 0.5 phase to the model curve for the primary's accretion rate in Figure~\ref{fig:havar}. The qualitative agreement between the model and the data suggests that our adjustments to the model, while crude, are capturing something real about the accretion geometry in the \eb\ system.

\begin{figure*}[ht]
    \centering
    \includegraphics[width=0.65\linewidth]{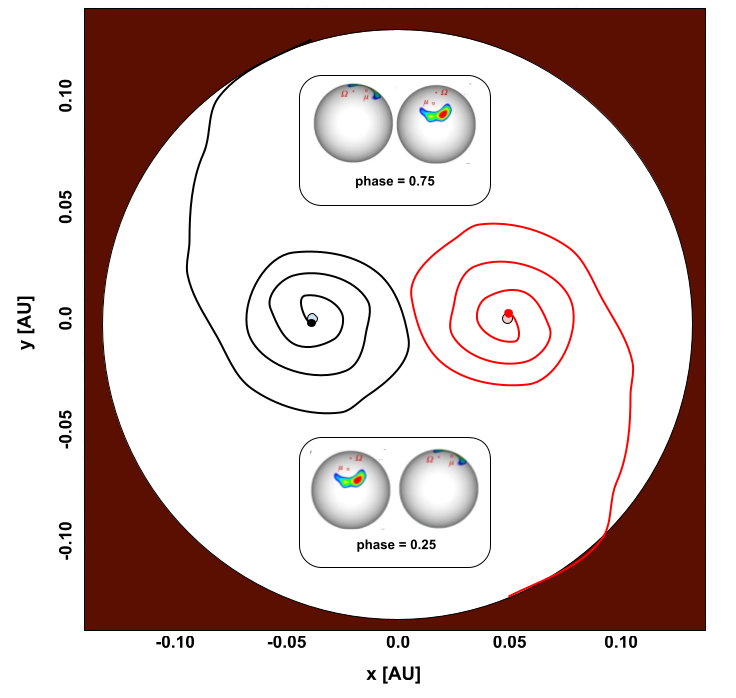}
    \caption{Schematic top-down representation of accretion onto the eclipsing stars in \eb\ via streams crossing the inner hole in the circumbinary disk. Small circles representing the two stars are to scale relative to their separation. The size of the cavity relative to the orbital semi-major axis is also to scale, inspired by the simulations of \citet{Artymowicz:1996}. Insets represent the accretion footpoints on the two stars as seen by the observer at the two quadrature phases \citep[taken from the modeling of][for the accretion footpoints observed in the young star GM~Aur]{Espaillat:2021}. The observer is in the direction of the bottom of the sketch.}
    \label{fig:streams}
\end{figure*}

In Figure~\ref{fig:streams} we offer a schematic depiction that may characterize the situation. The essential features of this conceptual sketch are the following: (1) the eclipsing binary sits in a cleared cavity or hole within a circumbinary disk that is presumably responsible for the infrared excess observed in the SED at 20~$\mu$m (see Figure~\ref{fig:sed}); (2) material from the inner disk falls in toward the central binary, with a cadence equal to the binary orbital period, along streams that are presumably optically thin and thus responsible for the emission in the SED at 10~$\mu$m (see Section~\ref{subsec:disk} below); (3) the streams produce hot accretion shocks on the stars' surfaces, again with a cadence equal to the binary orbit; and (4) these accretion hot spots are produced synchronously on the two stars but, due to the high inclination of the orbital plane to the line of sight and the synchronous rotation of the stars, only one star's spot is preferentially seen at one orbital quadrature and vice-versa. 
For an example of realistic accretion spots, we draw on recent work by \citet{Espaillat:2021} that mapped the detailed structure of the accretion footpoints on GM~Aur, a classical T~Tauri star; for conceptual simplicity, in Figure~\ref{fig:streams} we simply reproduce the GM~Aur accretion hot spot mapping for both stars but at opposite rotational phases. 
To be clear, we do not expect that these accretion footpoint ``spots" are the same as the (presumably dark) spots responsible for the rotational modulation discussed in Section~\ref{subsec:var}. That variability dominates the overall amplitude of variations in the TESS light curve and is coherent over the multi-year timescale of the TESS observations, whereas the accretion variability is highly stochastic and manifests itself principally in the H$\alpha$ full-width at 10\% intensity (see Figure~\ref{fig:havar}).

Finally, in light of the evidence for ongoing accretion in the system, it is reasonable to assume that the system experienced accretion in the past, and we may ask whether the accretion history could have affected the stellar properties \citep[see, e.g., the models of][]{Baraffe:2017,Vorobyov:2017}. 
%have investigated the effects of accretion ``bursts" on the internal and surface properties of young stars, for various assumptions 
\citet{Stassun:2014} considered whether these models could explain some of the discrepancies between standard model predictions and the observed properties of the available sample of PMS EBs, concluding that in general the predicted effects of accretion (i.e., undersized radii, increased \teff, under-luminosity, and enhanced Li depletion) were not consistent with the observed effects confronting many low-mass PMS EBs (i.e., inflated radii, suppressed \teff, and decreased Li depletion). In the present case, for example, the models of \citet{Baraffe:2017} predict for the \eb\ primary a radius of $\approx$0.88~\rsun\ (10\% smaller than observed), a \teff\ of $\approx$4350~K (600~K hotter than observed), and Li depletion $\sim$1.5~dex greater than observed (see Section~\ref{subsec:chromo}).

\subsection{Circumbinary disk and accretion streams}\label{subsec:disk}

The SED of \eb\ (Figure~\ref{fig:sed}) shows a $\gtrsim 7\sigma$ excess at 22~$\mu$m, suggesting the presence of warm dust at a blackbody temperature of a few hundred K. For example, the dotted green curve in Figure~\ref{fig:sed} represents a simple, passive dust disk model from \citet{Jensen:1997} with an 0.25~AU inner hole and temperature at the inner edge of $\sim$150~K. Such a model can reproduced the observed 22~$\mu$m excess well, however it does not reproduce the observed $\sim 6\sigma$ excess at 10$\mu$m, and scaling up the same disk model to match the 10$\mu$m excess would produce far too large an excess at 22~$\mu$m. 

Therefore, as was the case for the analogous system DQ~Tau \citep[see][]{Mathieu:1997}, we introduce a small amount of optically thin, hot dust within the hole of the circumbinary disk model. As shown by the solid green curve in Figure~\ref{fig:sed}, even just $8\times10^{-10}$~\msun\ (i.e., $3\times10^{-4}$~M$_\oplus$) of dust at 1500~K (corresponding to the dust sublimation temperature) produces a strong 10~$\mu$m silicate emission feature that matches the observed excess. It also produces a small amount of continuum emission in the near-IR that could explain the $2\sigma$ excess observed at 4.5~$\mu$m, though that excess may not be statistically significant. 

The need to include hot dust within the circumbinary disk is consistent with the idea that disk material is reaching the inner stars, which necessitates crossing the disk cavity, but not occupying a large cross-section of the cavity so as to produce emission at 10~$\mu$m but not over-produce emission at other wavelengths (see Section~\ref{subsec:accretion} and Figure~\ref{fig:streams}). 

Next, we investigated whether dust at the inner disk edge could potentially explain the residual dips observed in the TESS light curve after removal of the starspot signals (see Section~\ref{subsec:accretion} and Figure~\ref{fig:lcresid}). Based on the system geometry, in particular the inclination angle of the orbital plane (Table~\ref{tab:dimensions}) and assuming a disk cavity 0.25~AU in diameter (see above), we can calculate that the scale-height of the inner disk edge must be $h/r \approx 0.08$ (which corresponds to 2.4~\rsun) in order to allow disk material to ``transit" one of the central stars as seen by the observer. The dips appear to have a typical duration of $\sim$0.1 orbital phase ($\sim$0.3~d), and the Keplerian period at the inner disk edge is $\sim$26.5~d. As seen from the inner disk edge, the disc of one of the central stars subtends an angle of $\sim$3.6$^\circ$, thus material at the disk edge must travel 0.01 of its Keplerian orbit, or $\sim$0.27~d, to transit one of the stars. Thus, an explanation for the observed dips arising from transits by dust at the inner disk edge is fully consistent with the general picture that emerges for the \eb\ system (Figure~\ref{fig:streams}).
A similar interpretation has been proffered in other young EBs with circumbinary disks exhibiting similar behavior \citep[see, e.g.,][]{Terquem:2015}.

Finally, recent simulations of orbitally pulsed accretion suggest that the inner binary must possess one or more asymmetries in its physical characteristics (i.e., mass ratio or eccentricity) for the accretion bursts to occur on the orbital period, whereas for a fully symmetric binary ($e = 0$, $q = 1$) the accretion bursts recur every $\sim$5 orbital periods \citep[see, e.g.,][]{Munoz:2016}. In \eb, while $e=0$, the stars are evidently of sufficiently different mass ($q \approx 0.9$) to provide the symmetry breaking needed for orbitally pulsed accretion on the orbital period. The tendency for accretion rate variations in \eb\ to be greatest on the lower-mass secondary star (see Section~\ref{subsec:accretion}) is also consistent with the simulations \citep[e.g.,][]{Shi:2012,Munoz:2016}.

\subsection{Effects of magnetic activity on stellar properties}\label{subsec:chromo}

We found in Section~\ref{subsec:models} that the precisely measured radii and temperatures of the eclipsing stars in the \eb\ system are very well fit at the age of LCC by PMS evolutionary models that account for the effects on stellar structure by magnetic fields (Figure~\ref{fig:magnetic_models}). In contrast, the temperatures and radii cannot both be simultaneously fit by {\it standard} PMS evolutionary models (Figure~\ref{fig:std_models}) at any age (the models predict temperatures $\approx$200~K hotter than observed), and that the age implied by those standard models for the measured mass-radius relationship ($\approx$12~Myr) is significantly younger than the age of $16.2 \pm 2.2$~Myr for the LCC association to which \eb\ belongs (Section~\ref{subsec:age}). Put another way, at the age of LCC, the standard models imply that the stars in \eb\ are significantly cooler and larger than expected for their masses. 

Precisely this combination of effects---so-called radius inflation and temperature suppression---has been observed in other young, low-mass stars that are magnetically active. And there is ample evidence to suggest that the stars in \eb\ are indeed magnetically active, including their rapid rotation due to synchronization with the 3.1-day orbital period (Section~\ref{subsec:liha}), the presence of starspot modulations in the light curve (Section~\ref{subsec:var}), and the modulation of accretion diagnostics in a manner that suggests the presence of magnetic accretion footpoints on the stars (Section~\ref{subsec:accretion}). 

Based on the low-mass EB 2M0535$-$05 \citep{Stassun:2006,Stassun:2007} and a set of well characterized active M-dwarfs in the field, \citet{Stassun:2012} developed empirical relationships between a star's chromospheric activity, as measured by the strength of H$\alpha$ emission, and the degree to which the star's radius is inflated and its temperature suppressed. The physics underlying those empirical relationships has been suggested to be the reduction of surface flux emitted by a star with magnetically inhibited convective efficiency and/or covered by magnetic starspots, leading to a decreased overall effective temperature, in turn causing an enlarged radius so as to still radiate the luminosity produced in the stellar core \citep[see, e.g.,][]{Chabrier:2007,Somers:2017}.

\begin{figure}[!t]
%\epsscale{1.15}
\includegraphics[width=\linewidth,trim=0 0 95 180,clip]{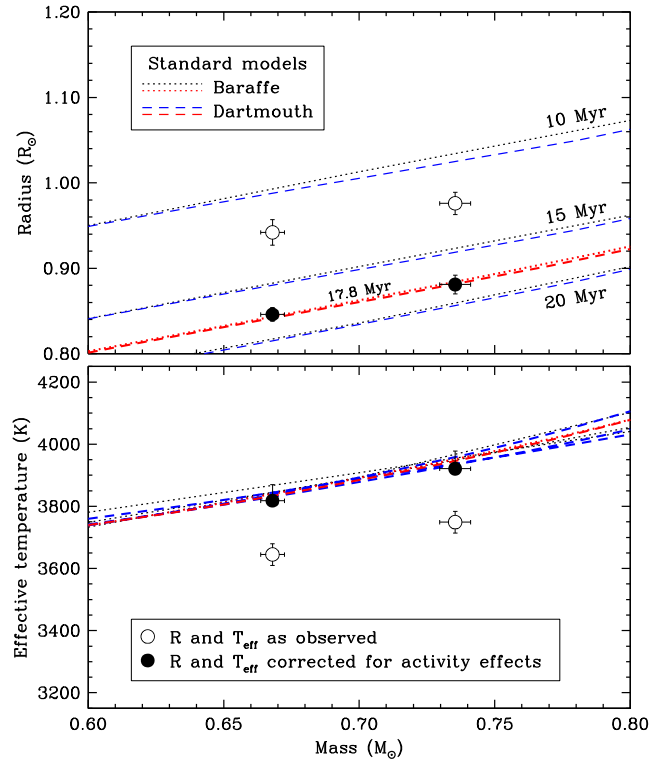}
\figcaption{Same as Figure~\ref{fig:std_models} (and on the same scale), now comparing the properties of \eb\ before (open symbols) and after (solid symbols) applying the chromospheric activity corrections of \citet{Stassun:2012} to the measured temperatures and radii. Now both the radii and temperatures of the two components are well matched by standard theory at an age of $\approx$17~Myr, the same age suggested by the magnetic models (Figure~\ref{fig:magnetic_models}). For clarity, the temperatures and radii represented by the solid symbols are {\it not} the ``true" values; rather, these are what the temperatures and radii {\it would be} if the stars were not magnetically active.\label{fig:standard_models_activ}}
\end{figure}

Figure~\ref{fig:standard_models_activ} shows the result of applying the \citet{Stassun:2012} relations to the measured temperatures and radii of the eclipsing stars in \eb. To apply the relations, we used the median H$\alpha$ equivalent widths as an estimate of the basal (presumably chromospheric) emission, to avoid instances of increased H$\alpha$ emission arising from episodic accretion (see Section~\ref{subsec:accretion} and Figure~\ref{fig:havar}). The agreement of the ``activity-corrected" temperatures and radii with the predictions of {\it standard} PMS evolutionary models---at an age consistent with the age of LCC---is remarkable, and lends strong support to the idea that strong stellar magnetic fields have significant and measurable effects on the fundamental properties of young stars and on their inferred ages. 

\begin{figure}[!t]
%\epsscale{1.15}
\includegraphics[width=\linewidth,trim=0 0 70 377,clip]{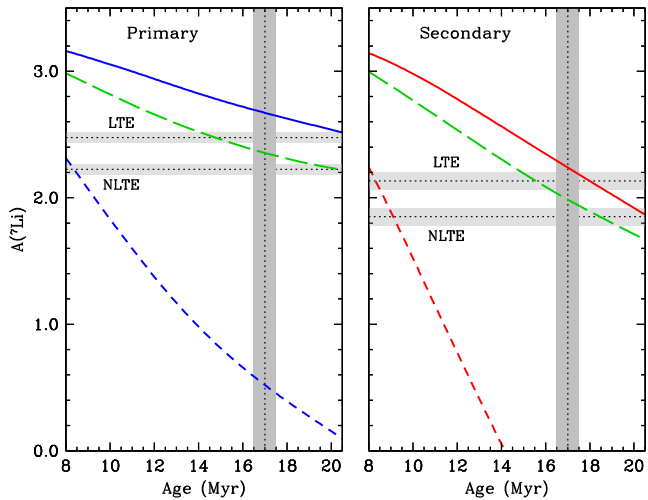}
\figcaption{Comparison of Li abundances measured for the primary (left) and secondary (right) eclipsing stars in the \eb\ system versus PMS evolutionary model predictions as a function of age; the adopted age of \eb\ is represented by the vertical swath. Measured abundances are represented as horizontal swaths for both LTE and NLTE cases (see Section~\ref{subsec:liha}). Evolutionary tracks for standard PMS models \citep[dashed curves;][]{Baraffe:2015} predict significantly more Li depletion than observed, by $\gtrsim$2~dex, whereas magnetic PMS models \citep[solid curves;][]{Feiden:2016} are able to much more nearly match the measured values. It is possible to match the Li abundances even better by adjusting the magnetic models with additional parameters, such as with different field strengths for the two stars, represented here by the SPOTS models \citep{Somers:2020} with spot coverage fractions of 24\% and 28\% for the primary and secondary, respectively (long-dashed green curves).\label{fig:li_models}}
\end{figure}

As a further test of these ideas, we also considered the Li abundances of the eclipsing stars in \eb\ in relation to the predictions of magnetic versus standard PMS evolutionary models. First, we converted the measured Li equivalent widths (Section~\ref{subsec:liha}) into abundances using the curves of growth from \citet{Pavlenko:1996}, which for the primary give $A({\rm Li})_1 = 2.48 \pm 0.04$~dex and $2.23 \pm 0.04$~dex for the LTE and NLTE cases, respectively; for the secondary, we obtain $A({\rm Li})_2 = 2.13 \pm 0.07$~dex and $1.85 \pm 0.07$~dex for the LTE and NLTE cases, respectively. These are represented in Figure~\ref{fig:li_models} in comparison to the same PMS evolutionary models considered above.

It is very clear that the standard models predict significantly more Li depletion at 17~Myr than observed, by $\gtrsim$2~dex. For those models, the observed Li abundances would imply a much younger age of $\sim$8~Myr. 
It is interesting that this age does not agree with the age of 12~Myr implied by these same models in the mass-radius diagram (see Figure~\ref{fig:std_models}). 
In addition, whereas the standard models are able to reproduce the ``activity-corrected" radii and temperatures at an age of 17~Myr (see Figure~\ref{fig:standard_models_activ}), the same cannot work with the Li abundances; adopting the warmer activity-corrected temperatures would make the observed Li abundances even larger, requiring an even younger inferred age by the standard models. 

Thus, while the standard models can predict the global properties (radii and temperatures) of magnetically active stars with suitable adjustments to stellar structure, those models cannot prevent Li from becoming rapidly depleted without an additional change to heat transport in the interior.
A similar effect was observed in the PMS EB V1174~Ori \citep{Stassun:2004}, which led those authors to conclude that convection in those magnetically active stars was much less efficient than in standard models \citep[see also][]{DAntona:2003}. 

In contrast, the magnetic models are able to reproduce the observed Li abundances reasonably well at an age of 17~Myr. An essential feature of these models is that the same magnetic effects by which the surface temperatures are suppressed and the radii inflated also result in less efficient convective heat transport and a decrease in the temperature at the base of the convection zone, and thus a greatly reduced rate of Li depletion \citep[see, e.g.,][]{Somers:2014,Somers:2015a,Somers:2015b,Somers:2020}. 

The fact that these models perhaps over-predict the Li abundances in \eb\ by 0.1--0.2~dex could suggest that a small modification is needed to the interiors physics of the models. For example, a small amount of convective overshoot at the base of the convection zone could in principle produce a small amount of additional Li depletion to match the observed abundances in this case. 
Alternatively, a slight reduction in the field strength assumed by the models could in principle allow the convective efficiency to be slightly increased and permit a small increase in the Li depletion. 
For example, the SPOTS magnetic models of \citet{Somers:2020} produce too much Li depletion with spot coverage fraction of 17\% and too little depletion with spot coverage fraction of 34\%, but can be made to match the measured values well with $\sim$25\% spot coverage (24\% on the primary, 28\% on the secondary; see Figure~\ref{fig:li_models}).

\subsection{The tertiary companion}\label{subsec:tertiary}

We conclude with some observations regarding the faint tertiary companion revealed by the speckle observations (Section~\ref{subsec:ao}). We have very few observational constraints available to characterize the object, other than a brightness contrast relative to the central EB at one wavelength and its contribution to the total light of the system required to be consistent with the third-light determined from the eclipse model (see Section~\ref{subsec:sed}). 

Its angular separation in the speckle imaging places it at a separation of $\sim$100~AU from the central EB, thus presumably it limits the circumbinary disk surrounding the EB to be no larger in extent than $\sim$100~AU. At the nominal 16.2~Myr age of the system, the PMS models of \citet{Baraffe:2015} imply a mass of $\sim$0.2~\msun, thus placing the object close to but likely above the substellar mass boundary.

More generally, it is interesting that the \eb\ system is a hierarchical triple. A large fraction of short-period PMS EBs are now known to be hierarchical triples \citep[see, e.g.,][]{Stassun:2014}, consistent with the very high prevalence of tertiaries observed among tight spectroscopic binaries in the field \citep[e.g.,][]{Tokovinin:2006,Laos:2020}.

\section{Summary and Conclusions}\label{sec:summary}

From TESS light curve observations, we have identified \eb\ as a low-mass, pre--main-sequence (PMS) eclipsing binary (EB) with a 3.07-day orbital period. Its membership in the Lower Centaurus Crux (LCC) association provides an opportunity to confront theoretical PMS evolutionary models with the added constraint of an independent age determination. Applying the neural-net age estimator of \citet{McBride:2021} to the {\it Gaia\/} astrometry and photometry, we infer an age for LCC and thereby for the \eb\ system of 16.2$\pm$2.2~Myr. 

There is evidence in the form of excess emission at $\gtrsim$10~$\mu$m for a circumbinary disk around the EB. In addition, speckle-imaging observations reveal a tertiary companion with a photometrically estimated mass of $\sim$0.2~\msun\ at $\sim$100~AU separation; the tertiary presumably limits the extent of the circumbinary disk to no more than 100~AU. \eb\ joins the large fraction of short-period PMS EBs that are hierarchical triples \citep[see, e.g.,][]{Stassun:2014}.

We observe periodic variations in the H$\alpha$ emission from both stars, in sync with the binary orbital period, which we interpret as accretion streams from the inner edge of the circumbinary disk to the central stars, consistent with theoretical predictions of dynamical interactions between binaries in circumbinary disks \citep[e.g.,][]{Artymowicz:1996} and previously reported in observations of some other PMS binaries \citep[e.g.,][]{Mathieu:1997,Jensen:2007}. Short-duration dips observed in the TESS light curve of \eb\ can also be explained by dust at the inner disk edge transiting the central stars, as has been reported for a number of other PMS stars with disks \citep[e.g.,][]{Terquem:2015,Rebull:2015,Stauffer:2015}.

A joint analysis of the TESS light curve data together with spectroscopically determined radial-velocity measurements and additional tight constraints imposed by the broadband spectral energy distribution (SED) and the {\it Gaia\/} parallax, we determine accurate, empirical masses, radii, and effective temperatures for both eclipsing stars in the system. The $\sim$1\% precision achieved on the masses and radii are among the best reported for PMS EBs \citep[see][for a review]{Stassun:2014}. Importantly, thanks to the extremely precise parallax supplied by {\it Gaia}, and following the precepts of \citet{Miller:2020}, we are able to achieve similarly good precision on the effective temperatures. This is important because traditionally it has been the systematic uncertainty on \teff\ \citep[typically $\sim$100~K; see][]{Torres:2010,StassunTorres:2016} that has been the limiting factor in the ability of EBs to fully stress-test models. 

Armed with these extremely precise measurements of both stars' masses, radii, temperatures, and independently estimated system age, we find that standard PMS stellar evolutionary models \citep[e.g.,][]{Baraffe:2015,Feiden:2016} are unable to simultaneously match the observed radii and temperatures at the age of LCC. Relative to the models, the stars are larger (radius-inflated) and cooler (temperature-suppressed) than predicted for their masses and age. Such effects have been reported among magnetically active low-mass PMS EB stars \citep[e.g.,][]{Stassun:2006,Stassun:2008}, and have been attributed to the reduction of surface flux emitted by a star with magnetically inhibited convective efficiency and/or covered by magnetic starspots, leading to a decreased overall effective temperature, in turn causing an enlarged radius so as to still radiate the luminosity produced in the stellar core \citep[see, e.g.,][]{Chabrier:2007,Somers:2017}. Indeed, the eclipsing stars in \eb\ are clearly magnetically active, as evinced by the presence of rotational starspot modulation signals in the light curve and strong H$\alpha$ emission from both stars in the spectra. 

We find that PMS evolutionary models that include the effects of surface magnetic fields \citep[e.g.,][]{Feiden:2016,Somers:2020} are able to reproduce the observed radii and temperatures in \eb\ with $\sim$1\% precision and at an age that is within 1~Myr of the nominal age of LCC. Moreover, applying the empirical ``activity corrections" of \citet{Stassun:2012} to the radii and temperatures based on the observed H$\alpha$ strengths---essentially making the stars appear as they would, were they not magnetically active---the {\it standard} PMS models are then able to successfully and precisely reproduce the radii and temperatures at the age of LCC. This strongly suggests that the standard models are ``correct" insofar as they represent stars without the effects of magnetic activity, and that the magnetic models are ``correct" in their implementation of those effects. 

These effects are also predicted to alter the interior structure of the stars, with strong implications for the rate at which elements such as Li are depleted. Indeed, we measure the Li abundances of the \eb\ stars to be $\sim$2~dex less depleted than predicted by the standard models, whereas the magnetic models are able to much more successfully reproduce the observed abundances. At a more detailed level, we find that the magnetic models of \citet{Feiden:2016} slightly over-predict the Li abundances compared to our measurements, possibly suggesting the need for some adjustment to the field strengths assumed in the models \citep[see, e.g.,][for an alternative implementation]{Somers:2020} or in some other aspect of the change in convective efficiency as incorporated in the models.

To a remarkable degree, the \eb\ system presents very strong evidence that magnetic activity in young stars alters both their global properties and the physics of their interiors.

%\vfill\eject 

\vspace{1in}
\acknowledgements
%\section*{ACKNOWLEDGMENTS}
K.G.S.\ acknowledges support from NASA ADAP grant 80NSSC20K0447.
%{The authors are grateful to the anonymous referee for their helpful critiques that helped to improve the manuscript.}
CHIRON observations were conducted at CTIO through time allocated by NSF's OIR Lab (NOIRLab Proposal ID 2020A-0146; PI: Bouma).
Other observations used the High-Resolution Imaging instrument Zorro, under Gemini LLP Proposal Number GN/S-2021A-LP-105. Zorro was funded by the NASA Exoplanet Exploration Program and built at the NASA Ames Research Center by S.B.\ Howell, N.\ Scott, E.P.\ Horch, and E.\ Quigley. 
%Zorro was mounted on the Gemini North (and/or South) telescope of the 
The international Gemini Observatory is a program of NSF's OIR Lab, managed by the Association of Universities for Research in Astronomy (AURA) under a cooperative agreement with NSF on behalf of the Gemini partnership: NSF (United States), National Research Council (Canada), Agencia Nacional de Investigaci\'on y Desarrollo (Chile), Ministerio de Ciencia, Tecnolog\'ia e Innovaci\'on (Argentina), Minist\'erio da Ciência, Tecnologia, Inovações e Comunicações (Brazil), and Korea Astronomy and Space Science Institute (Republic of Korea).
\\
%All the TESS data used can be found in MAST: \dataset[10.17909/t9-nmc8-f686]{http://dx.doi.org/10.17909/t9-nmc8-f686}, \dataset[10.17909/3y7c-wa45]{http://dx.doi.org/10.17909/3y7c-wa45}, \dataset[10.17909/fwdt-2x66]{http://dx.doi.org/10.17909/fwdt-2x66}.
All of the data presented in this paper were obtained from the Mikulski Archive for Space Telescopes (MAST) at the Space Telescope Science Institute. The specific observations analyzed can be accessed via \dataset[10.17909/t9-nmc8-f686]{https://doi.org/10.17909/t9-nmc8-f686}, \dataset[10.17909/3y7c-wa45]{https://doi.org/10.17909/3y7c-wa45},  \dataset[10.17909/fwdt-2x66]{https://doi.org/10.17909/fwdt-2x66}.

\bibliography{final.bib}

\begin{thebibliography}{}
\expandafter\ifx\csname natexlab\endcsname\relax\def\natexlab#1{#1}\fi
\providecommand{\url}[1]{\href{#1}{#1}}
\providecommand{\dodoi}[1]{doi:~\href{http://doi.org/#1}{\nolinkurl{#1}}}
\providecommand{\doeprint}[1]{\href{http://ascl.net/#1}{\nolinkurl{http://ascl.net/#1}}}
\providecommand{\doarXiv}[1]{\href{https://arxiv.org/abs/#1}{\nolinkurl{https://arxiv.org/abs/#1}}}

\bibitem[{{Andersen} {et~al.}(1980){Andersen}, {Clausen}, \&
  {Nordstrom}}]{Andersen:1980}
{Andersen}, J., {Clausen}, J.~V., \& {Nordstrom}, B. 1980, in Close Binary
  Stars: Observations and Interpretation, ed. M.~J. {Plavec}, D.~M. {Popper},
  \& R.~K. {Ulrich}, Vol.~88, 81--86

\bibitem[{{Ardila} {et~al.}(2015){Ardila}, {Jonhs-Krull}, {Herczeg}, {Mathieu},
  \& {Quijano-Vodniza}}]{Ardila:2015}
{Ardila}, D.~R., {Jonhs-Krull}, C., {Herczeg}, G.~J., {Mathieu}, R.~D., \&
  {Quijano-Vodniza}, A. 2015, \apj, 811, 131,
  \dodoi{10.1088/0004-637X/811/2/131}

\bibitem[{{Artymowicz} \& {Lubow}(1996)}]{Artymowicz:1996}
{Artymowicz}, P., \& {Lubow}, S.~H. 1996, \apjl, 467, L77,
  \dodoi{10.1086/310200}

\bibitem[{{Baraffe} {et~al.}(2017){Baraffe}, {Elbakyan}, {Vorobyov}, \&
  {Chabrier}}]{Baraffe:2017}
{Baraffe}, I., {Elbakyan}, V.~G., {Vorobyov}, E.~I., \& {Chabrier}, G. 2017,
  \aap, 597, A19, \dodoi{10.1051/0004-6361/201629303}

\bibitem[{{Baraffe} {et~al.}(2015){Baraffe}, {Homeier}, {Allard}, \&
  {Chabrier}}]{Baraffe:2015}
{Baraffe}, I., {Homeier}, D., {Allard}, F., \& {Chabrier}, G. 2015, \aap, 577,
  A42, \dodoi{10.1051/0004-6361/201425481}

\bibitem[{{Basri} {et~al.}(1997){Basri}, {Johns-Krull}, \&
  {Mathieu}}]{Basri:1997}
{Basri}, G., {Johns-Krull}, C.~M., \& {Mathieu}, R.~D. 1997, \aj, 114, 781,
  \dodoi{10.1086/118510}

\bibitem[{{Binnendijk}(1960)}]{Binnendijk:1960}
{Binnendijk}, L. 1960, {Properties of double stars; a survey of parallaxes and
  orbits.} (University of Pennsylvania Press)

\bibitem[{{Bouma} {et~al.}(2019){Bouma}, {Hartman}, {Bhatti}, {Winn}, \&
  {Bakos}}]{Bouma:2019}
{Bouma}, L.~G., {Hartman}, J.~D., {Bhatti}, W., {Winn}, J.~N., \& {Bakos},
  G.~{\'A}. 2019, \apjs, 245, 13, \dodoi{10.3847/1538-4365/ab4a7e}

\bibitem[{{Bowler} {et~al.}(2019){Bowler}, {Hinkley}, {Ziegler}, {Baranec},
  {Gizis}, {Law}, {Liu}, {Shah}, {Shkolnik}, {Riaz}, \& {Riddle}}]{Bowler:2019}
{Bowler}, B.~P., {Hinkley}, S., {Ziegler}, C., {et~al.} 2019, \apj, 877, 60,
  \dodoi{10.3847/1538-4357/ab1018}

\bibitem[{{Brown}(2021)}]{Brown:2021}
{Brown}, A. G.~A. 2021, \araa, 59, 59,
  \dodoi{10.1146/annurev-astro-112320-035628}

\bibitem[{{Chabrier} {et~al.}(2007){Chabrier}, {Gallardo}, \&
  {Baraffe}}]{Chabrier:2007}
{Chabrier}, G., {Gallardo}, J., \& {Baraffe}, I. 2007, \aap, 472, L17,
  \dodoi{10.1051/0004-6361:20077702}

\bibitem[{{Czekala} {et~al.}(2016){Czekala}, {Andrews}, {Torres}, {Jensen},
  {Stassun}, {Wilner}, \& {Latham}}]{Czekala:2016}
{Czekala}, I., {Andrews}, S.~M., {Torres}, G., {et~al.} 2016, \apj, 818, 156,
  \dodoi{10.3847/0004-637X/818/2/156}

\bibitem[{{D'Antona} \& {Montalb{\'a}n}(2003)}]{DAntona:2003}
{D'Antona}, F., \& {Montalb{\'a}n}, J. 2003, \aap, 412, 213,
  \dodoi{10.1051/0004-6361:20031410}

\bibitem[{{David} {et~al.}(2019){David}, {Hillenbrand}, {Gillen}, {Cody},
  {Howell}, {Isaacson}, \& {Livingston}}]{David:2019}
{David}, T.~J., {Hillenbrand}, L.~A., {Gillen}, E., {et~al.} 2019, \apj, 872,
  161, \dodoi{10.3847/1538-4357/aafe09}

\bibitem[{{Dotter} {et~al.}(2008){Dotter}, {Chaboyer}, {Jevremovi{\'c}},
  {Kostov}, {Baron}, \& {Ferguson}}]{Dotter:2008}
{Dotter}, A., {Chaboyer}, B., {Jevremovi{\'c}}, D., {et~al.} 2008, \apjs, 178,
  89, \dodoi{10.1086/589654}

\bibitem[{{Eker} {et~al.}(2020){Eker}, {Soydugan}, {Bilir}, {Bak{\i}{\c{s}}},
  {Ali{\c{c}}avu{\c{s}}}, {{\"O}zer}, {Aslan}, {Alpsoy}, \&
  {K{\"o}se}}]{Eker:2020}
{Eker}, Z., {Soydugan}, F., {Bilir}, S., {et~al.} 2020, \mnras, 496, 3887,
  \dodoi{10.1093/mnras/staa1659}

\bibitem[{{El-Badry} {et~al.}(2021){El-Badry}, {Rix}, \&
  {Heintz}}]{ElBadry:2021}
{El-Badry}, K., {Rix}, H.-W., \& {Heintz}, T.~M. 2021, \mnras, 506, 2269,
  \dodoi{10.1093/mnras/stab323}

\bibitem[{{Espaillat} {et~al.}(2021){Espaillat}, {Robinson}, {Romanova},
  {Thanathibodee}, {Wendeborn}, {Calvet}, {Reynolds}, \&
  {Muzerolle}}]{Espaillat:2021}
{Espaillat}, C.~C., {Robinson}, C.~E., {Romanova}, M.~M., {et~al.} 2021, \nat,
  597, 41, \dodoi{10.1038/s41586-021-03751-5}

\bibitem[{Etzel(1981)}]{Etzel:1981}
Etzel, P.~B. 1981, in Photometric and Spectroscopic Binary Systems, ed. E.~B.
  Carling \& Z.~Kopal (Dordrecht: Springer Netherlands), 111--120

\bibitem[{{Feiden}(2015)}]{Feiden:2015}
{Feiden}, G.~A. 2015, in Astronomical Society of the Pacific Conference Series,
  Vol. 496, Living Together: Planets, Host Stars and Binaries, ed. S.~M.
  {Rucinski}, G.~{Torres}, \& M.~{Zejda}, 137.
\newblock \doarXiv{1506.02715}

\bibitem[{{Feiden}(2016)}]{Feiden:2016}
{Feiden}, G.~A. 2016, \aap, 593, A99, \dodoi{10.1051/0004-6361/201527613}

\bibitem[{{Feiden} \& {Chaboyer}(2013)}]{Feiden:2013}
{Feiden}, G.~A., \& {Chaboyer}, B. 2013, \apj, 779, 183,
  \dodoi{10.1088/0004-637X/779/2/183}

\bibitem[{{Foreman-Mackey} {et~al.}(2013){Foreman-Mackey}, {Hogg}, {Lang}, \&
  {Goodman}}]{Foreman-Mackey:2013}
{Foreman-Mackey}, D., {Hogg}, D.~W., {Lang}, D., \& {Goodman}, J. 2013, \pasp,
  125, 306, \dodoi{10.1086/670067}

\bibitem[{{Gelman} \& {Rubin}(1992)}]{Gelman:1992}
{Gelman}, A., \& {Rubin}, D.~B. 1992, Statistical Science, 7, 457,
  \dodoi{10.1214/ss/1177011136}

\bibitem[{{Gilliland} {et~al.}(2010){Gilliland}, {Jenkins}, {Borucki},
  {Bryson}, {Caldwell}, {Clarke}, {Dotson}, {Haas}, {Hall}, {Klaus}, {Koch},
  {McCauliff}, {Quintana}, {Twicken}, \& {van Cleve}}]{Gilliland:2010}
{Gilliland}, R.~L., {Jenkins}, J.~M., {Borucki}, W.~J., {et~al.} 2010, \apjl,
  713, L160, \dodoi{10.1088/2041-8205/713/2/L160}

\bibitem[{{G{\'o}mez Maqueo Chew} {et~al.}(2012){G{\'o}mez Maqueo Chew},
  {Stassun}, {Pr{\v{s}}a}, {Stempels}, {Hebb}, {Barnes}, {Heller}, \&
  {Mathieu}}]{Gomez:2012}
{G{\'o}mez Maqueo Chew}, Y., {Stassun}, K.~G., {Pr{\v{s}}a}, A., {et~al.} 2012,
  \apj, 745, 58, \dodoi{10.1088/0004-637X/745/1/58}

\bibitem[{{Goodman} \& {Weare}(2010)}]{Goodman:2010}
{Goodman}, J., \& {Weare}, J. 2010, Communications in Applied Mathematics and
  Computational Science, 5, 65, \dodoi{10.2140/camcos.2010.5.65}

\bibitem[{{Gregory}(2005)}]{Gregory:2005}
{Gregory}, P.~C. 2005, \apj, 631, 1198, \dodoi{10.1086/432594}

\bibitem[{{Howell} {et~al.}(2011){Howell}, {Everett}, {Sherry}, {Horch}, \&
  {Ciardi}}]{Howell:2011}
{Howell}, S.~B., {Everett}, M.~E., {Sherry}, W., {Horch}, E., \& {Ciardi},
  D.~R. 2011, \aj, 142, 19, \dodoi{10.1088/0004-6256/142/1/19}

\bibitem[{{Husser} {et~al.}(2013){Husser}, {Wende-von Berg}, {Dreizler},
  {Homeier}, {Reiners}, {Barman}, \& {Hauschildt}}]{husser2013}
{Husser}, T.-O., {Wende-von Berg}, S., {Dreizler}, S., {et~al.} 2013, \aap,
  553, A6

\bibitem[{{Irwin} {et~al.}(2011){Irwin}, {Quinn}, {Berta}, {Latham}, {Torres},
  {Burke}, {Charbonneau}, {Dittmann}, {Esquerdo}, {Stefanik}, {Oksanen},
  {Buchhave}, {Nutzman}, {Berlind}, {Calkins}, \& {Falco}}]{Irwin:2011}
{Irwin}, J.~M., {Quinn}, S.~N., {Berta}, Z.~K., {et~al.} 2011, \apj, 742, 123,
  \dodoi{10.1088/0004-637X/742/2/123}

\bibitem[{{Jaehnig} {et~al.}(2019){Jaehnig}, {Somers}, \&
  {Stassun}}]{Jaehnig:2019}
{Jaehnig}, K., {Somers}, G., \& {Stassun}, K.~G. 2019, \apj, 879, 39,
  \dodoi{10.3847/1538-4357/ab21cf}

\bibitem[{{Jensen} {et~al.}(2007){Jensen}, {Dhital}, {Stassun}, {Patience},
  {Herbst}, {Walter}, {Simon}, \& {Basri}}]{Jensen:2007}
{Jensen}, E. L.~N., {Dhital}, S., {Stassun}, K.~G., {et~al.} 2007, \aj, 134,
  241, \dodoi{10.1086/518408}

\bibitem[{{Jensen} \& {Mathieu}(1997)}]{Jensen:1997}
{Jensen}, E. L.~N., \& {Mathieu}, R.~D. 1997, \aj, 114, 301,
  \dodoi{10.1086/118475}

\bibitem[{{Kipping}(2010)}]{Kipping:2010}
{Kipping}, D.~M. 2010, \mnras, 408, 1758,
  \dodoi{10.1111/j.1365-2966.2010.17242.x}

\bibitem[{{Kiraga}(2012)}]{Kiraga:2012}
{Kiraga}, M. 2012, \actaa, 62, 67.
\newblock \doarXiv{1204.3825}

\bibitem[{{Kounkel}(2022)}]{pyxcsao}
{Kounkel}, M. 2022, PyXCSAO, 0.2,  Zenodo, \dodoi{10.5281/zenodo.6998993}

\bibitem[{{Kounkel} \& {Covey}(2019)}]{Kounkel:2019}
{Kounkel}, M., \& {Covey}, K. 2019, \aj, 158, 122,
  \dodoi{10.3847/1538-3881/ab339a}

\bibitem[{{Kounkel} {et~al.}(2020){Kounkel}, {Covey}, \&
  {Stassun}}]{Kounkel:2020}
{Kounkel}, M., {Covey}, K., \& {Stassun}, K.~G. 2020, arXiv e-prints,
  arXiv:2004.07261.
\newblock \doarXiv{2004.07261}

\bibitem[{{Kurtz} \& {Mink}(1998)}]{rvsao}
{Kurtz}, M.~J., \& {Mink}, D.~J. 1998, \pasp, 110, 934

\bibitem[{{Laos} {et~al.}(2020){Laos}, {Stassun}, \& {Mathieu}}]{Laos:2020}
{Laos}, S., {Stassun}, K.~G., \& {Mathieu}, R.~D. 2020, \apj, 902, 107,
  \dodoi{10.3847/1538-4357/abb3fe}

\bibitem[{{Lindegren} {et~al.}(2018){Lindegren}, {Hern{\'a}ndez}, {Bombrun},
  {Klioner}, {Bastian}, {Ramos-Lerate}, {de Torres}, {Steidelm{\"u}ller},
  {Stephenson}, {Hobbs}, {Lammers}, {Biermann}, {Geyer}, {Hilger}, {Michalik},
  {Stampa}, {McMillan}, {Casta{\~n}eda}, {Clotet}, {Comoretto}, {Davidson},
  {Fabricius}, {Gracia}, {Hambly}, {Hutton}, {Mora}, {Portell}, {van Leeuwen},
  {Abbas}, {Abreu}, {Altmann}, {Andrei}, {Anglada}, {Balaguer-N{\'u}{\~n}ez},
  {Barache}, {Becciani}, {Bertone}, {Bianchi}, {Bouquillon}, {Bourda},
  {Br{\"u}semeister}, {Bucciarelli}, {Busonero}, {Buzzi}, {Cancelliere},
  {Carlucci}, {Charlot}, {Cheek}, {Crosta}, {Crowley}, {de Bruijne}, {de
  Felice}, {Drimmel}, {Esquej}, {Fienga}, {Fraile}, {Gai}, {Garralda},
  {Gonz{\'a}lez-Vidal}, {Guerra}, {Hauser}, {Hofmann}, {Holl}, {Jordan},
  {Lattanzi}, {Lenhardt}, {Liao}, {Licata}, {Lister}, {L{\"o}ffler},
  {Marchant}, {Martin-Fleitas}, {Messineo}, {Mignard}, {Morbidelli}, {Poggio},
  {Riva}, {Rowell}, {Salguero}, {Sarasso}, {Sciacca}, {Siddiqui}, {Smart},
  {Spagna}, {Steele}, {Taris}, {Torra}, {van Elteren}, {van Reeven}, \&
  {Vecchiato}}]{Lindegren:2018}
{Lindegren}, L., {Hern{\'a}ndez}, J., {Bombrun}, A., {et~al.} 2018, \aap, 616,
  A2, \dodoi{10.1051/0004-6361/201832727}

\bibitem[{{Lindegren} {et~al.}(2021){Lindegren}, {Bastian}, {Biermann},
  {Bombrun}, {de Torres}, {Gerlach}, {Geyer}, {Hern{\'a}ndez}, {Hilger},
  {Hobbs}, {Klioner}, {Lammers}, {McMillan}, {Ramos-Lerate},
  {Steidelm{\"u}ller}, {Stephenson}, \& {van Leeuwen}}]{Lindegren:2021}
{Lindegren}, L., {Bastian}, U., {Biermann}, M., {et~al.} 2021, \aap, 649, A4,
  \dodoi{10.1051/0004-6361/202039653}

\bibitem[{{Luri} {et~al.}(2018){Luri}, {Brown}, {Sarro}, {Arenou},
  {Bailer-Jones}, {Castro-Ginard}, {de Bruijne}, {Prusti}, {Babusiaux}, \&
  {Delgado}}]{Luri:2018}
{Luri}, X., {Brown}, A.~G.~A., {Sarro}, L.~M., {et~al.} 2018, \aap, 616, A9,
  \dodoi{10.1051/0004-6361/201832964}

\bibitem[{{MacDonald} \& {Mullan}(2014)}]{MacDonald:2014}
{MacDonald}, J., \& {Mullan}, D.~J. 2014, \apj, 787, 70,
  \dodoi{10.1088/0004-637X/787/1/70}

\bibitem[{{Mathieu} {et~al.}(1997){Mathieu}, {Stassun}, {Basri}, {Jensen},
  {Johns-Krull}, {Valenti}, \& {Hartmann}}]{Mathieu:1997}
{Mathieu}, R.~D., {Stassun}, K., {Basri}, G., {et~al.} 1997, \aj, 113, 1841,
  \dodoi{10.1086/118395}

\bibitem[{McBride {et~al.}(2021)McBride, Lingg, Kounkel, Covey, \&
  Hutchinson}]{McBride:2021}
McBride, A., Lingg, R., Kounkel, M., Covey, K., \& Hutchinson, B. 2021, The
  Astronomical Journal, 162, 282, \dodoi{10.3847/1538-3881/ac2432}

\bibitem[{{Miller} {et~al.}(2020){Miller}, {Maxted}, \&
  {Smalley}}]{Miller:2020}
{Miller}, N.~J., {Maxted}, P.~F.~L., \& {Smalley}, B. 2020, \mnras, 497, 2899,
  \dodoi{10.1093/mnras/staa2167}

\bibitem[{{Mu{\~n}oz} \& {Lai}(2016)}]{Munoz:2016}
{Mu{\~n}oz}, D.~J., \& {Lai}, D. 2016, \apj, 827, 43,
  \dodoi{10.3847/0004-637X/827/1/43}

\bibitem[{{Murphy} {et~al.}(2020){Murphy}, {Lawson}, {Onken}, {Yong}, {Da
  Costa}, {Zhou}, {Mamajek}, {Bell}, {Bessell}, \& {Feinstein}}]{Murphy:2020}
{Murphy}, S.~J., {Lawson}, W.~A., {Onken}, C.~A., {et~al.} 2020, \mnras, 491,
  4902, \dodoi{10.1093/mnras/stz3198}

\bibitem[{{Muzerolle} {et~al.}(2019){Muzerolle}, {Flaherty}, {Balog}, {Beck},
  \& {Gutermuth}}]{Muzerolle:2019}
{Muzerolle}, J., {Flaherty}, K., {Balog}, Z., {Beck}, T., \& {Gutermuth}, R.
  2019, \apj, 877, 29, \dodoi{10.3847/1538-4357/ab1756}

\bibitem[{{Oelkers} \& {Stassun}(2018)}]{Oelkers:2018}
{Oelkers}, R.~J., \& {Stassun}, K.~G. 2018, \aj, 156, 132,
  \dodoi{10.3847/1538-3881/aad68e}

\bibitem[{{Pavlenko} \& {Magazzu}(1996)}]{Pavlenko:1996}
{Pavlenko}, Y.~V., \& {Magazzu}, A. 1996, \aap, 311, 961.
\newblock \doarXiv{astro-ph/9606090}

\bibitem[{{Popper} \& {Etzel}(1981)}]{Popper:1981}
{Popper}, D.~M., \& {Etzel}, P.~B. 1981, \aj, 86, 102, \dodoi{10.1086/112862}

\bibitem[{{Preibisch} \& {Mamajek}(2008)}]{Preibisch:2008}
{Preibisch}, T., \& {Mamajek}, E. 2008, in Handbook of Star Forming Regions,
  Volume II, ed. B.~{Reipurth}, Vol.~5 (ASP Monograph Publications), 235

\bibitem[{{Pr{\v{s}}a} {et~al.}(2016){Pr{\v{s}}a}, {Harmanec}, {Torres},
  {Mamajek}, {Asplund}, {Capitaine}, {Christensen-Dalsgaard}, {Depagne},
  {Haberreiter}, {Hekker}, {Hilton}, {Kopp}, {Kostov}, {Kurtz}, {Laskar},
  {Mason}, {Milone}, {Montgomery}, {Richards}, {Schmutz}, {Schou}, \&
  {Stewart}}]{Prsa:2016}
{Pr{\v{s}}a}, A., {Harmanec}, P., {Torres}, G., {et~al.} 2016, \aj, 152, 41,
  \dodoi{10.3847/0004-6256/152/2/41}

\bibitem[{{Rebull} {et~al.}(2015){Rebull}, {Stauffer}, {Cody}, {G{\"u}nther},
  {Hillenbrand}, {Poppenhaeger}, {Wolk}, {Hora}, {Hernandez}, {Bayo}, {Covey},
  {Forbrich}, {Gutermuth}, {Morales-Calder{\'o}n}, {Plavchan}, {Song}, {Bouy},
  {Terebey}, {Cuillandre}, \& {Allen}}]{Rebull:2015}
{Rebull}, L.~M., {Stauffer}, J.~R., {Cody}, A.~M., {et~al.} 2015, \aj, 150,
  175, \dodoi{10.1088/0004-6256/150/6/175}

\bibitem[{{Ricker} {et~al.}(2015){Ricker}, {Winn}, {Vanderspek}, {Latham},
  {Bakos}, {Bean}, {Berta-Thompson}, {Brown}, {Buchhave}, {Butler}, {Butler},
  {Chaplin}, {Charbonneau}, {Christensen-Dalsgaard}, {Clampin}, {Deming},
  {Doty}, {De Lee}, {Dressing}, {Dunham}, {Endl}, {Fressin}, {Ge}, {Henning},
  {Holman}, {Howard}, {Ida}, {Jenkins}, {Jernigan}, {Johnson}, {Kaltenegger},
  {Kawai}, {Kjeldsen}, {Laughlin}, {Levine}, {Lin}, {Lissauer}, {MacQueen},
  {Marcy}, {McCullough}, {Morton}, {Narita}, {Paegert}, {Palle}, {Pepe},
  {Pepper}, {Quirrenbach}, {Rinehart}, {Sasselov}, {Sato}, {Seager},
  {Sozzetti}, {Stassun}, {Sullivan}, {Szentgyorgyi}, {Torres}, {Udry}, \&
  {Villasenor}}]{Ricker:2015}
{Ricker}, G.~R., {Winn}, J.~N., {Vanderspek}, R., {et~al.} 2015, Journal of
  Astronomical Telescopes, Instruments, and Systems, 1, 014003,
  \dodoi{10.1117/1.JATIS.1.1.014003}

\bibitem[{{Scott} {et~al.}(2021){Scott}, {Howell}, {Gnilka}, {Stephens},
  {Salinas}, {Matson}, {Furlan}, {Horch}, {Everett}, {Ciardi}, {Mills}, \&
  {Quigley}}]{Scott:2021}
{Scott}, N.~J., {Howell}, S.~B., {Gnilka}, C.~L., {et~al.} 2021, Frontiers in
  Astronomy and Space Sciences, 8, 138, \dodoi{10.3389/fspas.2021.716560}

\bibitem[{{Shi} {et~al.}(2012){Shi}, {Krolik}, {Lubow}, \& {Hawley}}]{Shi:2012}
{Shi}, J.-M., {Krolik}, J.~H., {Lubow}, S.~H., \& {Hawley}, J.~F. 2012, \apj,
  749, 118, \dodoi{10.1088/0004-637X/749/2/118}

\bibitem[{{Somers} {et~al.}(2020){Somers}, {Cao}, \&
  {Pinsonneault}}]{Somers:2020}
{Somers}, G., {Cao}, L., \& {Pinsonneault}, M.~H. 2020, \apj, 891, 29,
  \dodoi{10.3847/1538-4357/ab722e}

\bibitem[{{Somers} \& {Pinsonneault}(2014)}]{Somers:2014}
{Somers}, G., \& {Pinsonneault}, M.~H. 2014, \apj, 790, 72,
  \dodoi{10.1088/0004-637X/790/1/72}

\bibitem[{{Somers} \& {Pinsonneault}(2015{\natexlab{a}})}]{Somers:2015a}
---. 2015{\natexlab{a}}, \apj, 807, 174, \dodoi{10.1088/0004-637X/807/2/174}

\bibitem[{{Somers} \& {Pinsonneault}(2015{\natexlab{b}})}]{Somers:2015b}
---. 2015{\natexlab{b}}, \mnras, 449, 4131, \dodoi{10.1093/mnras/stv630}

\bibitem[{{Somers} \& {Stassun}(2017)}]{Somers:2017}
{Somers}, G., \& {Stassun}, K.~G. 2017, \aj, 153, 101,
  \dodoi{10.3847/1538-3881/153/3/101}

\bibitem[{{Stassun} {et~al.}(2014){Stassun}, {Feiden}, \&
  {Torres}}]{Stassun:2014}
{Stassun}, K.~G., {Feiden}, G.~A., \& {Torres}, G. 2014, \nar, 60, 1,
  \dodoi{10.1016/j.newar.2014.06.001}

\bibitem[{{Stassun} {et~al.}(2012){Stassun}, {Kratter}, {Scholz}, \&
  {Dupuy}}]{Stassun:2012}
{Stassun}, K.~G., {Kratter}, K.~M., {Scholz}, A., \& {Dupuy}, T.~J. 2012, \apj,
  756, 47, \dodoi{10.1088/0004-637X/756/1/47}

\bibitem[{{Stassun} {et~al.}(2008){Stassun}, {Mathieu}, {Cargile}, {Aarnio},
  {Stempels}, \& {Geller}}]{Stassun:2008}
{Stassun}, K.~G., {Mathieu}, R.~D., {Cargile}, P.~A., {et~al.} 2008, \nat, 453,
  1079, \dodoi{10.1038/nature07069}

\bibitem[{{Stassun} {et~al.}(1999){Stassun}, {Mathieu}, {Mazeh}, \&
  {Vrba}}]{Stassun:1999}
{Stassun}, K.~G., {Mathieu}, R.~D., {Mazeh}, T., \& {Vrba}, F.~J. 1999, \aj,
  117, 2941, \dodoi{10.1086/300881}

\bibitem[{{Stassun} {et~al.}(2006){Stassun}, {Mathieu}, \&
  {Valenti}}]{Stassun:2006}
{Stassun}, K.~G., {Mathieu}, R.~D., \& {Valenti}, J.~A. 2006, \nat, 440, 311,
  \dodoi{10.1038/nature04570}

\bibitem[{{Stassun} {et~al.}(2007){Stassun}, {Mathieu}, \&
  {Valenti}}]{Stassun:2007}
---. 2007, \apj, 664, 1154, \dodoi{10.1086/519231}

\bibitem[{{Stassun} {et~al.}(2004){Stassun}, {Mathieu}, {Vaz}, {Stroud}, \&
  {Vrba}}]{Stassun:2004}
{Stassun}, K.~G., {Mathieu}, R.~D., {Vaz}, L. P.~R., {Stroud}, N., \& {Vrba},
  F.~J. 2004, \apjs, 151, 357, \dodoi{10.1086/382353}

\bibitem[{{Stassun} \& {Torres}(2016)}]{StassunTorres:2016}
{Stassun}, K.~G., \& {Torres}, G. 2016, \aj, 152, 180,
  \dodoi{10.3847/0004-6256/152/6/180}

\bibitem[{{Stassun} \& {Torres}(2021)}]{StassunTorres:2021}
---. 2021, \apjl, 907, L33, \dodoi{10.3847/2041-8213/abdaad}

\bibitem[{{Stassun} {et~al.}(2019){Stassun}, {Oelkers}, {Paegert}, {Torres},
  {Pepper}, {De Lee}, {Collins}, {Latham}, {Muirhead}, {Chittidi},
  {Rojas-Ayala}, {Fleming}, {Rose}, {Tenenbaum}, {Ting}, {Kane}, {Barclay},
  {Bean}, {Brassuer}, {Charbonneau}, {Ge}, {Lissauer}, {Mann}, {McLean},
  {Mullally}, {Narita}, {Plavchan}, {Ricker}, {Sasselov}, {Seager}, {Sharma},
  {Shiao}, {Sozzetti}, {Stello}, {Vanderspek}, {Wallace}, \&
  {Winn}}]{StassunTIC:2019}
{Stassun}, K.~G., {Oelkers}, R.~J., {Paegert}, M., {et~al.} 2019, \aj, 158,
  138, \dodoi{10.3847/1538-3881/ab3467}

\bibitem[{{Stauffer} {et~al.}(2014){Stauffer}, {Cody}, {Baglin}, {Alencar},
  {Rebull}, {Hillenbrand}, {Venuti}, {Turner}, {Carpenter}, {Plavchan},
  {Findeisen}, {Carey}, {Terebey}, {Morales-Calder{\'o}n}, {Bouvier}, {Micela},
  {Flaccomio}, {Song}, {Gutermuth}, {Hartmann}, {Calvet}, {Whitney}, {Barrado},
  {Vrba}, {Covey}, {Herbst}, {Furesz}, {Aigrain}, \& {Favata}}]{Stauffer:2014}
{Stauffer}, J., {Cody}, A.~M., {Baglin}, A., {et~al.} 2014, \aj, 147, 83,
  \dodoi{10.1088/0004-6256/147/4/83}

\bibitem[{{Stauffer} {et~al.}(2015){Stauffer}, {Cody}, {McGinnis}, {Rebull},
  {Hillenbrand}, {Turner}, {Carpenter}, {Plavchan}, {Carey}, {Terebey},
  {Morales-Calder{\'o}n}, {Alencar}, {Bouvier}, {Venuti}, {Hartmann}, {Calvet},
  {Micela}, {Flaccomio}, {Song}, {Gutermuth}, {Barrado}, {Vrba}, {Covey},
  {Padgett}, {Herbst}, {Gillen}, {Lyra}, {Medeiros Guimaraes}, {Bouy}, \&
  {Favata}}]{Stauffer:2015}
{Stauffer}, J., {Cody}, A.~M., {McGinnis}, P., {et~al.} 2015, \aj, 149, 130,
  \dodoi{10.1088/0004-6256/149/4/130}

\bibitem[{{Stauffer} {et~al.}(2017){Stauffer}, {Collier Cameron}, {Jardine},
  {David}, {Rebull}, {Cody}, {Hillenbrand}, {Barrado}, {Wolk}, {Davenport}, \&
  {Pinsonneault}}]{Stauffer:2017}
{Stauffer}, J., {Collier Cameron}, A., {Jardine}, M., {et~al.} 2017, \aj, 153,
  152, \dodoi{10.3847/1538-3881/aa5eb9}

\bibitem[{{Stauffer} {et~al.}(2021){Stauffer}, {Rebull}, {Jardine}, {Collier
  Cameron}, {Cody}, {Hillenbrand}, {Barrado}, {Kruse}, \&
  {Powell}}]{Stauffer:2021}
{Stauffer}, J., {Rebull}, L.~M., {Jardine}, M., {et~al.} 2021, \aj, 161, 60,
  \dodoi{10.3847/1538-3881/abc7c6}

\bibitem[{{Terquem} {et~al.}(2015){Terquem}, {S{\o}rensen-Clark}, \&
  {Bouvier}}]{Terquem:2015}
{Terquem}, C., {S{\o}rensen-Clark}, P.~M., \& {Bouvier}, J. 2015, \mnras, 454,
  3472, \dodoi{10.1093/mnras/stv2258}

\bibitem[{{Tofflemire} {et~al.}(2017{\natexlab{a}}){Tofflemire}, {Mathieu},
  {Ardila}, {Akeson}, {Ciardi}, {Johns-Krull}, {Herczeg}, \&
  {Quijano-Vodniza}}]{Tofflemire:2017dq}
{Tofflemire}, B.~M., {Mathieu}, R.~D., {Ardila}, D.~R., {et~al.}
  2017{\natexlab{a}}, \apj, 835, 8, \dodoi{10.3847/1538-4357/835/1/8}

\bibitem[{{Tofflemire} {et~al.}(2017{\natexlab{b}}){Tofflemire}, {Mathieu},
  {Herczeg}, {Akeson}, \& {Ciardi}}]{Tofflemire:2017twa}
{Tofflemire}, B.~M., {Mathieu}, R.~D., {Herczeg}, G.~J., {Akeson}, R.~L., \&
  {Ciardi}, D.~R. 2017{\natexlab{b}}, \apjl, 842, L12,
  \dodoi{10.3847/2041-8213/aa75cb}

\bibitem[{{Tofflemire} {et~al.}(2019){Tofflemire}, {Mathieu}, \&
  {Johns-Krull}}]{Tofflemire:2019}
{Tofflemire}, B.~M., {Mathieu}, R.~D., \& {Johns-Krull}, C.~M. 2019, \aj, 158,
  245, \dodoi{10.3847/1538-3881/ab4f7d}

\bibitem[{{Tofflemire} {et~al.}(2022){Tofflemire}, {Kraus}, {Mann}, {Newton},
  {Gully-Santiago}, {Vanderburg}, {Waalkes}, {Berta-Thompson}, {Collins},
  {Collins}, {Nielsen}, {Bouchy}, {Ziegler}, {Briceno}, \&
  {Law}}]{Tofflemire:2022}
{Tofflemire}, B.~M., {Kraus}, A.~L., {Mann}, A.~W., {et~al.} 2022, arXiv
  e-prints, arXiv:2210.10789.
\newblock \doarXiv{2210.10789}

\bibitem[{{Tokovinin} {et~al.}(2006){Tokovinin}, {Thomas}, {Sterzik}, \&
  {Udry}}]{Tokovinin:2006}
{Tokovinin}, A., {Thomas}, S., {Sterzik}, M., \& {Udry}, S. 2006, \aap, 450,
  681, \dodoi{10.1051/0004-6361:20054427}

\bibitem[{{Torres}(2013)}]{Torres:2013}
{Torres}, G. 2013, Astronomische Nachrichten, 334, 4,
  \dodoi{10.1002/asna.201211743}

\bibitem[{{Torres} {et~al.}(2010){Torres}, {Andersen}, \&
  {Gim{\'e}nez}}]{Torres:2010}
{Torres}, G., {Andersen}, J., \& {Gim{\'e}nez}, A. 2010, ARA\&A, 18, 67,
  \dodoi{10.1007/s00159-009-0025-1}

\bibitem[{{Vorobyov} {et~al.}(2017){Vorobyov}, {Elbakyan}, {Hosokawa},
  {Sakurai}, {Guedel}, \& {Yorke}}]{Vorobyov:2017}
{Vorobyov}, E.~I., {Elbakyan}, V., {Hosokawa}, T., {et~al.} 2017, \aap, 605,
  A77, \dodoi{10.1051/0004-6361/201630356}

\bibitem[{{White} \& {Basri}(2003)}]{White:2003}
{White}, R.~J., \& {Basri}, G. 2003, \apj, 582, 1109, \dodoi{10.1086/344673}

\end{thebibliography}

% \begin{thebibliography}{}

% % \bibitem[Arnal et al.(1988)]{Arnal:1988} Arnal, M., Morrell, N., Garcia, B., et al.\ 1988, \pasp, 100,
% % 1076

% % \bibitem[Claret \& Bloemen(2011)]{Claret:2011} Claret, A., \& Bloemen, S.\ 2011, \aap, 529, A75 

% % \bibitem[Etzel(1981)]{Etzel:1981} Etzel, P.\ B. 1981, Photometric and Spectroscopic Binary Systems,
% % Proc.\ NATO Adv.\ Study Inst., ed.\ E.\ B.\ Carling \& Z.\ Kopal (Dordrecht: Reidel), p.\ 111

% % \bibitem[Foreman-Mackey et al.(2013)]{Foreman-Mackey:2013} Foreman-Mackey, D., Hogg, D.\ W., Lang, D.,
% % \& Goodman, J.\ 2013, \pasp, 125, 306

% % \bibitem[Gelman \& Rubin(1992)]{Gelman:1992} Gelman, A., \& Rubin, D.\ B. 1992, Statistical Science, 7,
% % 457
% % \bibitem[Gilliland et al.(2010)]{Gilliland:2010} Gilliland, R.\ L., Jenkins, J.\ M., Borucki, W.\ J.,
% % et al.\ 2010, \apjl, 713, L160

% % \bibitem[Gregory(2005)]{Gregory:2005} Gregory, P.\ C.\ 2005, \apj, 631, 1198

% % \bibitem[Huang \& Gies(2006)]{Huang:2006} Huang, W. \& Gies, D.~R.\ 2006, \apj, 648, 591

% % \bibitem[Irwin et al.(2011)]{Irwin:2011} Irwin, J.\ M., Quinn, S.\ N., Berta, Z.\ K., et al.\ 2011,
% % \apj, 742, 123      

% % \bibitem[Kipping(2010)]{Kipping:2010} Kipping, D.\ M.\ 2010, \mnras, 408, 1758

% % \bibitem[Popper \& Etzel(1981)]{Popper:1981} Popper, D.\ M., \& Etzel, P.\ B. 1981, \aj, 86, 102

% % \bibitem[Scargle(1998)]{Scargle:1998} Scargle, J.~D.\ 1998, \apj, 504, 405. doi:10.1086/306064

% \bibitem[Wolk et al.(2008)]{Wolk:2008} Wolk, S.~J., Spitzbart, B.~D., Bourke, T.~L., et al.\ 2008, \aj, 135, 693. doi:10.1088/0004-6256/135/2/693

% \end{thebibliography}
\end{document}